\documentclass[aps,prd,twocolumn,superscriptaddress,nofootinbib]{revtex4-2}

\usepackage{amsfonts}
\usepackage{amssymb,amsmath,enumitem}
\usepackage{graphicx,color,epstopdf}
\usepackage{dcolumn}
\usepackage{bm}
\usepackage{xcolor}
\usepackage{subcaption}
\usepackage{hyperref}
\usepackage{orcidlink}
\usepackage{placeins}

\def\nn{\nonumber \\}
\def\l{\left}
\def\r{\right}

\begin{document}

\title{Systematic biases in gravitational-wave parameter estimation from neglecting orbital eccentricity in space-based detectors}

\author{Jin-Zhao Yang\orcidlink{0000-0002-4826-6014}}
\author{Jia-Hao Zhong\orcidlink{0009-0001-3101-3774}}
\author{Tao Yang\orcidlink{0000-0002-2161-0495}}
\email{Corresponding author: yangtao@whu.edu.cn}
\affiliation{School of Physics and Technology, Wuhan University, Wuhan 430072, China}

\date{\today}

\begin{abstract}
Accurate modeling of gravitational-wave signals is essential for reliable inference of compact-binary source parameters, particularly for future space-based detectors operating in the milli- and deci-Hertz bands. In this work, we systematically investigate the parameter-estimation biases induced by neglecting orbital eccentricity when analyzing eccentric compact-binary coalescences with quasicircular waveform templates. Focusing on the deci-Hertz detector B-DECIGO and the milli-Hertz detector LISA, we model eccentric inspiral signals using a frequency-domain waveform that incorporates eccentricity-induced higher harmonics and the time-dependent response of spaceborne detectors. We quantify systematic biases in the chirp mass, symmetric mass ratio, and luminosity distance using both Bayesian inference and the Fisher-Cutler-Vallisneri (FCV) formalism, and assess their significance relative to statistical uncertainties. By constructing mock gravitational-wave catalogs spanning stellar-mass and massive black-hole binaries, we identify critical initial eccentricities at which systematic errors become comparable to statistical errors. We find that, for B-DECIGO, adopting a one-year observation period with $f_{\text{min}} = 0.1$ Hz, even very small eccentricities, $e_0 \sim 10^{-4}-10^{-3}$ at 0.1 Hz, can lead to significant biases, whereas for LISA, adopting a five-year observation period with $f_{\text{min}} = 10^{-4}$ Hz, such effects typically arise at larger eccentricities, $e_0 \sim 10^{-2}-10^{-1}$ at $10^{-4}$ Hz, due to the smaller number of in-band cycles. Comparisons between FCV predictions and full Bayesian analyses show good agreement over most of the initial eccentricity range, but diverge at high eccentricity, where Bayesian inference is required. Our results highlight the necessity of incorporating eccentricity in waveform models for future space-based gravitational-wave observations.
\end{abstract}

\maketitle

\section{Introduction} \label{sec:intro}

The detection of gravitational waves (GWs) has inaugurated a golden era in astrophysics, cosmology, and the exploration of strong-field gravity. With the current generation of ground-based detectors operating as a global network, the LIGO-Virgo-KAGRA (LVK) collaboration \cite{LIGOScientific:2014pky, VIRGO:2014yos, Somiya:2011np} has reported more than 200 GW events in the recently released Gravitational Wave Transient Catalog 4.0 (GWTC-4.0) \cite{LIGOScientific:2025hdt}. The majority of these detections originate from stellar-mass binary black hole (BBH) mergers, with a few arising from binary neutron star (BNS) and neutron star–black hole (NSBH) systems. Collectively, BBH, BNS, and NSBH mergers constitute the three subclasses of compact binary coalescences (CBCs), which represent the most promising sources of detectable GWs.
GW observations \cite{LIGOScientific:2018mvr, LIGOScientific:2020ibl, KAGRA:2021vkt, LIGOScientific:2021usb, LIGOScientific:2025hdt} have enabled significant advances in precision astrophysics, cosmology, and fundamental physics. These include studies of astrophysical formation channels and merger rates \cite{KAGRA:2021duu, LIGOScientific:2025pvj}, constraints on dark matter \cite{LIGOScientific:2021ffg}, cosmological measurements \cite{LIGOScientific:2021aug, LIGOScientific:2025jau}, and tests of general relativity (GR) in the strong-field regime \cite{LIGOScientific:2019fpa, LIGOScientific:2021sio}.

Accurate and computationally efficient GW waveform templates are essential for extracting astrophysical information from observed signals. The matched-filter technique allows for the extraction of GW signals from background noise by using a physically correct waveform template of the expected signal as the filter \cite{Allen:2005fk}. Parameter estimation (PE) stands as a central step in GW data analysis, enabling the inference of source properties from detected signals. Bayesian inference has become the standard and most robust framework for PE \cite{Christensen:2022bxb}, relying on the generation of a large number of theoretical waveforms to explore the parameter space and identify the posterior distribution \cite{Thrane:2018qnx}. As GW catalog expands and detector techniques improve both in sensitivities and frequency band coverage, waveform models of increasingly high accuracy are required to explore finer structures in the likelihood landscape. Physical characteristics previously considered subdominant, such as orbital eccentricity, higher-order multipole modes, and spin-induced precession, are expected to become increasingly relevant. Failure to properly account for these effects in the waveform may result in systematic errors that limit the efficiency and precision of GW extraction and inference.

This paper focuses on quantifying the systematic biases introduced by neglecting orbital eccentricity in waveform templates. Eccentricity characterizes the noncircularity of a binary orbit and provides a powerful diagnostic for distinguishing between different astrophysical formation channels of CBCs \cite{Zevin:2021rtf, Dorozsmai:2025jlu}. For binary merger formed in isolation, its own GW emission efficiently circularizes the orbit, with the eccentricity evolving approximately at a rate of $e_t/e_0 = (f_t/f_0)^{-19/18}$, where $e_t$ is the eccentricity at GW frequency $f_t$ and $e_0$ is defined at a reference frequency $f_0$ \cite{Peters:1964zz}. As a result, binaries formed through isolated evolution are expected to exhibit negligible eccentricity by the time their signals enter the LVK frequency band. On the other hand, dynamically formed binaries may retain measurable in-band eccentricities due to interactions with external bodies \cite{Zevin:2021rtf, Rodriguez:2017pec, Rodriguez:2018jqu, Tagawa:2020jnc}. Such scenarios include formation in dense stellar environments, such as globular clusters, young or open clusters, and nuclear star clusters (please refer to \cite{DallAmico:2023neb, Dorozsmai:2025jlu} and references therein), as well as interactions with gaseous discs of active galactic nuclei \cite{Yang:2019cbr, McKernan:2019beu, Ishibashi:2020zzy, Samsing:2020tda}, central supermassive black holes \cite{Tagawa:2020jnc, Bartos:2016dgn, Yang:2020xyi, Gayathri:2021xwb, Camilloni:2023xvf, Cocco:2025adu, Cocco:2025udb}, and hierarchical mergers \cite{Kimball:2020qyd, Antonini:2024het, Li:2025fnf, MaganaHernandez:2025fkm, Antonini:2025ilj, Tong:2025wpz, Cocco:2025adu, Cocco:2025udb}. The presence of eccentricity can also improve PE accuracy by breaking degeneracies among intrinsic and extrinsic parameters. Previous studies have reported notable improvements in sky localization \cite{Yang:2022fgp, Yang:2022tig, Yang:2022iwn}, while the additional harmonic structure induced by eccentricity enables earlier detection and more precise localization of incoming signals \cite{Yang:2023zxk, Yang:2024vfy}, it also leads to tighter cosmological constraints \cite{Zhong:2026pjg}.

Several studies have reported evidence for eccentric GW events using eccentric waveform templates \cite{Romero-Shaw:2020thy, Gayathri:2020coq, Gamba:2021gap, Romero-Shaw:2022xko, Gupte:2024jfe, Morras:2025xfu, Planas:2025jny, Planas:2025plq, Romero-Shaw:2025vbc, McMillin:2025hof}, although the regular inclusion of eccentricity in PE pipelines remains an open issue. Analyzing eccentric signals with quasicircular templates can introduce significant systematic biases, and much effort has been put on this topic. Cho \textit{et al}. \cite{Cho:2022cdy} investigated eccentricity-induced biases for BNS signals based on LIGO using both Bayesian inference and the Fisher-Cutler-Vallisneri (FCV) method. This study was extended to the LIGO-Virgo network by Divyajyoti \textit{et al}. \cite{Divyajyoti:2023rht}, and subsequently to third-generation detectors, Einstein Telescope (ET) \cite{ET:2019dnz, ET:2025xjr} and Cosmic Explorer (CE) \cite{Evans:2021gyd}, by Roy \textit{et al}. \cite{DuttaRoy:2024aew}, with a focus on the bias in tidal deformability. Recently, Das \textit{et al}. demonstrated that neglecting eccentricity can lead to substantial biases in the LVK network \cite{Das:2024zib}, and Divyajyoti \textit{et al}. showed that an eccentric aligned-spin model may be preferred over a quasicircular precessing-spin model \cite{Divyajyoti:2025cwq}. Huez \textit{et al}. further reported noticeable biases for moderately eccentric signals in synthetic BNS catalogs \cite{Huez:2025npe}.
More recently, RoyChowdhury \textit{et al}. studied PE biases arising from unmodeled eccentricity in the LVK network, comparing results obtained using recovery templates with different spin configurations \cite{RoyChowdhury:2026xgb}.
Omitting eccentricity can also bias tests of GR, potentially leading to spurious claims of deviations from GR. The impact of eccentricity omission on parametrized tests of gravity has been explored by Saini \textit{et al}. \cite{Saini:2022igm, Saini:2023rto}, Narayan \textit{et al}. \cite{Narayan:2023vhm}, and Roy \textit{et al}. \cite{Roy:2025xih}. Alternatively, the impact on inspiral-merger-ringdown consistency tests have been investigated by Bhat \textit{et al}. \cite{Bhat:2022amc} and Shaikh \textit{et al}. \cite{Shaikh:2024wyn}. These studies emphasized the risk of misinterpretation when eccentric effects are ignored.

With the advent of future GW detectors, the accessible detecting band will further extend. Space-based detectors such as DECIGO \cite{Isoyama:2018rjb, Kawamura:2020pcg} and LISA \cite{LISA:2024hlh}, operating in the decihertz and millihertz bands respectively, are expected to observe stellar-mass BBHs or massive black hole binaries with substantially higher signal-to-noise ratios (SNRs). At lower frequencies, GW signals retain measurable residual eccentricity with a higher possibility, and the accuracy of PE will be susceptible to the omission of eccentricity. 
Nevertheless, systematic investigations based on space-based detectors remain limited. Initial studies by Choi \textit{et al}. \cite{GilChoi:2022waq} examined eccentricity-induced biases for decihertz detectors B-DECIGO and MAGIS, while Garg \textit{et al}. \cite{Garg:2024oeu, Garg:2024qxq} explored the systematics in estimation of binary parameters and in tests of GR based on LISA.
What is more, eccentricity modifies GW waveforms not only through corrections to amplitude and phase, but also by introducing additional harmonic content \cite{Yunes:2009yz, Huerta:2014eca, Islam:2025bhf}. These harmonics may lead to further systematic effects if omitted, particularly for space-based detectors capable of tracking GW signals over year-long duty periods.
In this work, we investigate the impact of eccentricity on parameter estimation with considering response of additional harmonics on spaceborne detectors, and focus on the decihertz detector B-DECIGO and the millihertz detector LISA. Using a mock GW catalog with sources distributed uniformly on sky, we employ both FCV and Bayesian inference methods to systematically quantify the PE biases arising from neglecting eccentricity. We assess the significance of these systematic errors by comparing them with the corresponding statistical uncertainties.

The paper is organized as follows. In Sec. \ref{sec:waveform} we introduce the modeling of eccentric waveforms on space-based detectors, including the role of eccentricity-induced harmonics and the incorporation of detector motion in the frequency-domain response. Section \ref{sec:methods} describes the Bayesian and FCV methods used to quantify systematic biases. In Sec. \ref{sec:results}, we present the mock GW catalog and analyze the dependence of systematic errors on the omitted eccentricity. Section \ref{sec:conclusion} summarizes our findings and discusses their implications. Technical details and derivations are provided in the Appendixes.

\section{Modeling the eccentric response on space-based detectors} \label{sec:waveform}

\subsection{Eccentric waveform with harmonic structure}

In recent years, much effort has been devoted to the development of frequency-domain eccentric waveform models suitable for upcoming observing runs, as well as to assessing the detectability of eccentricity and its impact on parameter estimation \cite{Konigsdorffer:2006zt, Yunes:2009yz, Klein:2010ti, Mishra:2015bqa, Moore:2016qxz, Tanay:2016zog, Klein:2018ybm, Boetzel:2019nfw, Moore:2019xkm, Tiwari:2019jtz, Klein:2021jtd, Khalil:2021txt, Paul:2022xfy, Henry:2023tka, Morras:2025nlp, Phurailatpam:2025ppf}. These inspiral-only waveform models are constructed to achieve sufficient accuracy for comparison with numerical relativity (NR) simulations, in the meantime retaining sufficient computational efficiency for direct application in PE analyses.
In this work, we adopt the nonspinning, inspiral-only EccentricFD waveform approximant, implemented in \textsc{LALSuite} \cite{lalsuite} and generated using \textsc{PyCBC} \cite{Biwer:2018osg}, to efficiently model GW signals with low to moderate eccentricities. The EccentricFD approximant corresponds to the enhanced postcircular (EPC) model introduced in \cite{Huerta:2014eca}, which satisfies the following key requirements:
(i) at the zeroth order in eccentricity, it recovers the 3.5 post-Newtonian (PN) TaylorF2 waveform \cite{Buonanno:2009zt};
(ii) at the leading PN order, it reduces to the postcircular (PC) expansion \cite{Yunes:2009yz}, with eccentricity corrections up to $\mathcal{O}(e^8)$.

The EccentricFD waveform depends on 11 parameters
\begin{equation}
\vec\lambda = (\mathcal M_c, \eta, d_L, \iota, \theta, \phi, \psi, t_c, \phi_c, e_0, \beta)\,,
\end{equation}
where $\mathcal M_c = \mu^{3/5} M^{2/5}$ is the chirp mass, with $M = m_1 + m_2$ denoting the total mass and $\mu = m_1 m_2 / M$ the reduced mass; $\eta = \mu / M$ is the symmetric mass ratio; $d_L$ is the luminosity distance; $(\theta, \phi)$ specify the sky location of the source; $\psi$ is the polarization angle; $(t_c, \phi_c)$ are the coalescence time and phase; and $(e_0, \beta)$ are additional eccentric parameters compared with the TaylorF2 approximant. $e_0$ denotes the orbital eccentricity defined at a reference frequency $f_0$, while $\beta$ represents the azimuthal component of the inclination angle, which becomes relevant for non-circular orbits. 
The full frequency-domain waveform includes harmonic contributions induced by eccentricity, with harmonic indices ranging from $\ell = 1$ to $10$ \cite{Huerta:2014eca},
\begin{equation}
    \label{eccfdfunc}
    \tilde{h}(f) = -\sqrt{\frac{5}{384} \frac{\mathcal{M}_c^{5/6}}{\pi^{2/3} d_L}} f^{-7/6} \sum_{\ell=1}^{10} \xi_\ell \left(\frac{\ell}{2}\right)^{2/3} e^{-i\Psi_\ell}\,,
\end{equation}
and the phase associated with each harmonic is given by
\begin{equation}
    \label{eq:placeholder_label}
    \Psi_\ell = 2\pi f t_c - \ell \phi_c + \left(\frac{\ell}{2}\right)^{8/3} \frac{3}{128 \eta v_{\text{ecc}}^5} \sum_{n=0}^{7} a_n v_{\text{ecc}}^n\,,
\end{equation}
where $\xi_\ell$ is a function of $(\iota, \theta, \phi, \psi, \beta)$ that encodes the eccentricity-dependent amplitude modulation, $v_{\text{ecc}}(f; e_0)$ is a modified velocity parameter that smoothly interpolates between the postcircular expansion and the 3.5 PN TaylorF2 limit, and $a_n$ are the PN phase coefficients up to $n=7$. Explicit expressions for $\xi_\ell$, $v_{\text{ecc}}$, and the PN coefficients can be found in \cite{Huerta:2014eca}.
In the limit $e_0 \rightarrow 0$, the waveform reduces to the quasicircular TaylorF2 model, with only the dominant quadrupole harmonic ($\ell = 2$) present. As the eccentricity increases, additional harmonics become increasingly relevant, introducing richer oscillatory structure in the phase of the signal. Each harmonic is associated with a characteristic frequency list related to the dominant quadrupole mode by $f_\ell = \ell f_2 / 2 = \ell F$, where $F$ denotes the orbital frequency. With eccentricity corrections up to $\mathcal{O}(e^8)$ and harmonic indices up to $\ell = 10$, the validity of the EccentricFD model is therefore restricted to moderate eccentricities, $e_0 \in (0,0.4)$.

\subsection{Constructing the evolutionary frequency-domain space-based detector response}

\begin{figure}
    \centering
    \includegraphics[width=0.9\linewidth]{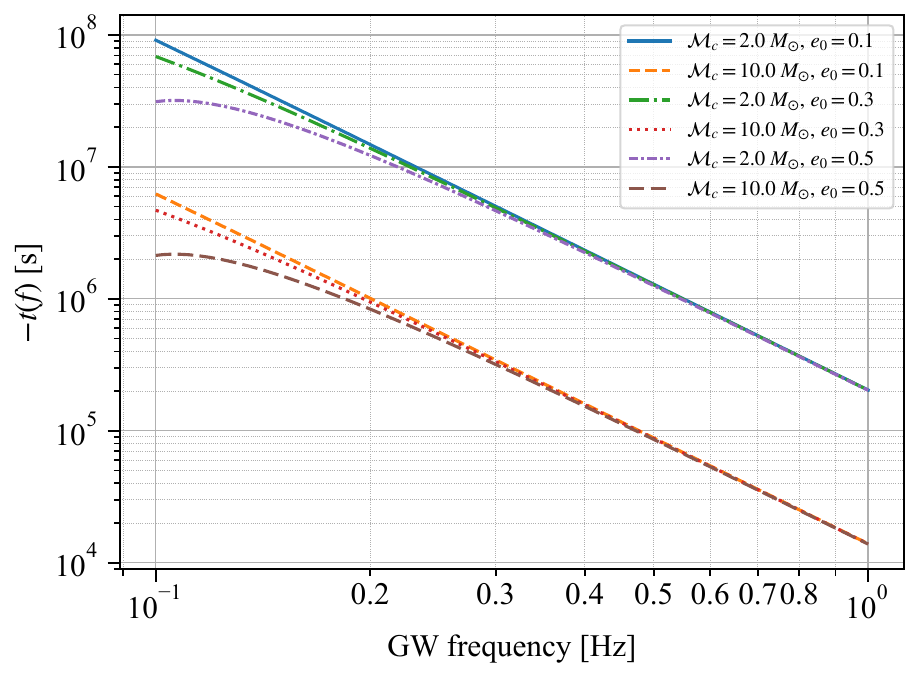}
    \caption{Example eccentric time-frequency relations for different values of $\mathcal M_c$ and $e_0$, with $e_0$ defined at $f_0 = 0.1$ Hz}
    \label{tf}
\end{figure}

In the mid- and lower-frequency bands, GW signals typically exhibit long inspiral durations, ranging from hours to several years. As a result, the orbital motion of the detector within the Solar System cannot be neglected. The detector response to an incoming GW signal can be written as
\begin{equation}
    h(t) = F_+(t) h_+(t) + F_\times(t) h_{\times}(t)\,,
\end{equation}
where $F_{+,\times}(t)$ are the time-dependent detector antenna pattern functions. For spaceborne detectors, these response functions vary slowly over the course of the observation, typically on timescales of a year.
This slow temporal modulation allows the detector response to be approximately reformulated in the frequency domain. The polarized frequency-domain waveforms can be obtained either by Fourier-transforming the time-domain signals or by applying the stationary phase approximation, yielding
\begin{equation}
    \tilde h(f) = F_+\big(t(f)\big) \tilde h_+(f) + F_\times\big(t(f)\big) \tilde h_{\times}(f)\,.
\end{equation}
Here, the time–frequency mapping $t(f)$ relates the GW frequency to the corresponding detector time along the inspiral.
We use the quadrupole-mode frequency $f_2 = 2F$ as the baseline frequency to construct the time-frequency relation $t(f)$. Following treatments for spaceborne detectors in \cite{Rubbo:2003ap, Cutler:1997ta}, the detector response functions $F_{+,\times}\big(t(f)\big)$ are computed in terms of the quadrupole frequency. The time-frequency relation $t(f)$ is obtained by solving the phase evolution equation (Eq.~(4.24)), together with Eqs.~(2.5) and (3.11) in \cite{Yunes:2009yz}. For each eccentric harmonic $\ell$, the corresponding detector response shall be evaluated at $F_{+,\times}(t(2f_\ell / \ell))$. This procedure is fully consistent with the eccentricity-induced structure of the EccentricFD waveform model. An example of the resulting eccentric time-frequency relation for different values of the chirp mass $\mathcal M_c$ and initial eccentricity $e_0$, defined at $f_0 = 0.1$ Hz, is shown in Fig.~\ref{tf}. As illustrated, binaries with larger initial eccentricities evolve more rapidly, leading to shorter inspiral durations within the detector’s sensitive band.

In this study, we consider two spaceborne detectors: B-DECIGO, a pathfinder for DECIGO, and LISA. For B-DECIGO, we adopt an arm length of $L = 100$ km and an observation time of 1 year, while for LISA we use $L = 2.5 \times 10^6$ km and an observation time of 5 years. The corresponding detector frequency ranges are $f_{\text{min}} = 0.1$ Hz and $f_{\text{max}} = 10$ Hz for B-DECIGO, and $f_{\text{min}} = 10^{-4}$ Hz and $f_{\text{max}} = 0.1$ Hz for LISA.
The starting frequency $f_{\text{start}}$ of the quadrupole mode is determined by the condition
\begin{equation} \label{cal_fstart}
    t_{\text{upper}} - t(f_{\text{start}}) = 
    \begin{cases}
    1~\text{year}, & \text{B-DECIGO}\,, \\
    5~\text{years}, & \text{\ \ \ \ LISA}\,,
    \end{cases}
\end{equation}
where $t_{\text{upper}}$ corresponds to the earlier of the following two times: the time associated with the upper bound of the detector’s frequency band, or the time at which the binary reaches the innermost stable circular orbit (ISCO).
For each eccentric harmonic $\ell$, the corresponding starting frequency is given by $\ell f_{\text{start}} / 2$. The waveform is therefore truncated according to
\begin{equation}
    \tilde h_{\text{obs}} (f) = \tilde h(f) \, \mathcal H\left(2f - \ell f_{\text{start}}\right)\,,
\end{equation}
where $\mathcal H(x)$ denotes the Heaviside step function,
\begin{equation}
    \mathcal{H}(x) = 
    \begin{cases}
        1, & x \geq 0\,, \\
        0, & x < 0\,.
    \end{cases}
\end{equation}

As an inspiral-only waveform approximant, EccentricFD terminates the signal at the ISCO frequency,
\begin{equation}
    f_{\text{ISCO}} = \frac{1}{6^{3/2} \pi M} \simeq 220 \left(\frac{M}{20 M_\odot}\right)^{-1} \text{Hz}\,.
\end{equation}
Ultimately, the effective frequency range used in the GW data analysis is determined jointly by the detector’s frequency band, the starting frequency $f_{\text{start}}$ and the ISCO frequency $f_{\text{ISCO}}$.

\section{Statistical Methods to Quantify the Systematics} \label{sec:methods}

\subsection{Bayesian inference}

Using the one-sided noise power spectral density (PSD), the PSD-weighted inner product between two signals $a$ and $b$ is defined as
\begin{equation}
    (a|b) = 4 \int_{f_{\text{min}}}^{f_{\text{max}}} \frac{\tilde a^\ast(f)\tilde b(f) + \tilde a(f)\tilde b^\ast(f)}{2 \text{Sn}(f)} df\,.
\end{equation}
The sensitivity curve $\text{Sn}(f)$ for B-DECIGO is adopted from \cite{Yagi:2011wg} and rescaled following \cite{Kawamura:2020pcg}, and for LISA we use the sensitivity model presented in \cite{LISACosmologyWorkingGroup:2019mwx}. The notation $(\cdot|\cdot)$ is reserved for the PSD-weighted inner product throughout this work, unless stated otherwise.

We assume that the detector noise is stationary, Gaussian, and statistically independent between frequency bins. Under this assumption, the Gaussian noise likelihood for GW data in a single frequency bin $j$, given a set of waveform parameters $\vec\lambda$, can be written as \cite{Thrane:2018qnx}
\begin{eqnarray}
    \mathcal{L}(s_j  |  \vec\lambda) &=& \frac{1}{\sqrt{2\pi \text{Sn}(f_j)}} \cdot \nn
    && \exp \bigg( -2 \delta f \cdot \frac{| 
    s_j - \tilde h_j(\vec\lambda) |^2}{\text{Sn}(f_j)} \bigg )\,,
\end{eqnarray}
where $\delta f$ is the frequency resolution, $s_j$ denotes the observed data in the $j$th frequency bin, and $\tilde h_j(\vec\lambda)$ is the template waveform evaluated at $f_j$. We emphasize that $\mathcal L(s_j | \vec\lambda)$ denotes the conditional likelihood given $\vec \lambda$ and $(s_j | \vec\lambda)$ in this context does not represent an inner product.
The total likelihood for the full dataset ${s_j}$ is given by the product
\begin{equation}
    \mathcal L(\{s_j\} |  \vec \lambda) = \prod_j \mathcal L(s_j | \vec \lambda)\,.
\end{equation}
It is often convenient to work with the logarithm of the likelihood, which converts the product into a sum. The log-likelihood can be expressed as a summation
\begin{equation}
    \log \mathcal{L}(s  |  \vec\lambda) = \mathcal N - \frac{1}{2} \l( s - \tilde h(\vec\lambda) \Big| s - \tilde h(\vec\lambda) \r)\,,
\end{equation}
where the constant $\mathcal N$ is
\begin{equation}
    \mathcal N = -\sum_j \log(2\pi \text{Sn}(f_j)) / 2\,.
\end{equation}
For a network consisting of $N$ detectors, the combined likelihood is obtained by multiplying the individual detector likelihoods $\mathcal L = \prod_n \mathcal L_n$.

Applying Bayes’s theorem, the posterior probability distribution for the parameter set $\vec\lambda$ can be written as
\begin{equation}
    p(\vec\lambda | s) = \frac{\mathcal L(s |  \vec\lambda) \pi(\vec\lambda)}{\mathcal Z}\,,
\end{equation}
where $\pi(\vec\lambda)$ denotes the prior distribution. $\mathcal Z$ is the Bayesian evidence
\begin{equation}
    \mathcal Z = \int \mathcal L(s |  \vec\lambda) \pi(\vec\lambda) d\vec\lambda\,.
\end{equation}
The maximum \textit{a posteriori} (MAP) estimate corresponds to the point in parameter space that maximizes the posterior density. 

Systematic bias is quantified as
\begin{equation}
    \Delta \vec\lambda^{\text{sys}} = |\vec \lambda^{\text{AP}} - \vec \lambda^{\text{inj}}|\,,
\end{equation}
where $\vec\lambda^{\text{AP}}$ denotes the estimated parameters obtained using a waveform template that omits eccentricity effects, and $\vec\lambda^{\text{inj}}$ are the injected values. Statistical uncertainties are assessed using credible intervals derived from the marginal posterior distributions. In particular, the $1\sigma$ uncertainty is defined as the interval enclosing 68\% of the posterior probability, bounded by the corresponding lower and upper limits.

\subsection{Fisher information matrix and Fisher-Cutler-Vallisneri method}

To theoretically estimate and quantify the systematics, we adopt the linear-signal approximation (LSA), and assume Gaussian distributions for waveform parameters. Statistical uncertainties are evaluated using the Fisher information matrix (FIM) \cite{Cutler:1994ys}:
\begin{equation}
    \Gamma_{ij} = \l( \frac{\partial h}{\partial \lambda_i} \bigg | \frac{\partial h}{\partial \lambda_j} \r)\,,
\end{equation}
where $\lambda_i$ denotes one of the 11 waveform parameters. In the quasicircular limit, only 9 parameters are relevant, with $e_0$ and $\beta$ excluded. The covariance matrix is given by $C_{ij} = \Gamma_{ij}^{-1}$, from which the marginalized $1\sigma$ uncertainties are obtained as $\sigma_i = \sqrt{C_{ii}}$. This method agrees well with Bayesian estimates for signals with sufficiently high SNRs.

Systematic biases arise when the waveform template used in parameter estimation differs from the true GW signal. By denoting such errors as $\Delta \vec\lambda^{\text{sys}}$, the Fisher-Cutler-Vallisneri (FCV) method expresses them as \cite{Cutler:2007mi, Chandramouli:2024vhw, Dhani:2024jja}
\begin{eqnarray}
    \label{fcv1}
    \Delta \lambda^{\text{sys}}_{j} &=& C^{ij}|_{h^{\text{AP}}(\vec\lambda^{\text{tr}})} \cdot \nn
    && \Big(\partial_{\lambda_i} h^{\text{AP}}(\vec\lambda^{\text{tr}}) \Big| h^s (\vec\lambda^{\text{tr}}) - h^{\text{AP}}(\vec\lambda^{\text{tr}}) \Big)\,,
\end{eqnarray}
where the linear approximation of the waveform is assumed \cite{Cutler:2007mi, Chandramouli:2024vhw}:
\begin{equation}\label{wave_lexp}
    h^{\text{AP}}(\vec\lambda^{\text{tr}}) \simeq h^{\text{AP}}(\vec\lambda^{\text{bf}}) - \delta \lambda_i \partial_{\lambda_i} h^{\text{AP}}(\vec\lambda^{\text{tr}})\,.
\end{equation}
Here, $h^s$ denotes the true GW signal (eccentric, with $e_0 \neq 0$), while $h^{\text{AP}}$ represents the waveform approximant used in the data analysis. The parameter shifts are defined as $\delta \lambda_i = \lambda^{\text{bf}}_i - \lambda^{\text{tr}}_i$, where $\lambda^{\text{bf}}_i$ are the best-fit parameters obtained using $h^{\text{AP}}$, and $\lambda^{\text{tr}}_i = \lambda^{\text{inj}}_i$ denote the true injected parameters.
To investigate the impact of neglecting eccentricity, we construct $h^{\text{AP}}$ by setting $e_0 = 0$, so that EccentricFD reduces to TaylorF2 waveform. Extensions to a detector network are described in Appendix \ref{appxfcv}.

The FCV method provides a direct estimate of $\Delta \vec\lambda^{\text{sys}}$. For clarity and easy comparison, in the next section we report only its absolute value. To quantify the significance of the systematic bias for a given parameter, we normalize it by the corresponding statistical error $\sigma_i$. We define a critical initial eccentricity $e_0^{\text{cr}}$ as the value of $e_0$ at which $|\Delta \lambda^{\text{sys}}_i|$ exceeds the boundary of the 90\% credible interval (90\% CL) assuming a Gaussian posterior distribution or equivalently the ratio $|\Delta \lambda^{\text{sys}}_i / \sigma_i|$ first exceeds roughly 1.645. Here $|\cdot|$ represents the absolute value.

\subsection{Mismatch and waveform alignment}

As can be seen from Eq.~(\ref{fcv1}), the systematic bias is directly proportional to the waveform difference. Even for signals with identical intrinsic parameters, the difference can be significant due to variations in extrinsic parameters such as the coalescence time $t_c$ and phase $\phi_c$. By optimizing over these extrinsic parameters, the waveform mismatch can be minimized, ensuring the validity of the LSA and thereby improving the consistency between Bayesian and FCV methods of the bias estimation \cite{Damour:2010zb, Hu:2022rjq, Dhani:2024jja}.

The mismatch between two waveforms, $h_1$ and $h_2$, is defined as
\begin{equation}
    \label{mm}
    \mathcal M = \min_{t_c,\phi_c}\bigg\{1 - \frac{(h_1|h_2)}{\sqrt{(h_1|h_1)(h_2|h_2)}} \bigg\}\,.
\end{equation}
By minimizing Eq.~(\ref{mm}) over $t_c$ and $\phi_c$, the waveform yields a set of aligned parameters $\vec\lambda^{\text{align}}$ with updated $t_c$ and $\phi_c$, and Eq.~(\ref{fcv1}) becomes
\begin{eqnarray}
    \label{fcv2}
    \Delta \lambda^{\text{sys}}_{j} &=& C^{ij}|_{h^{\text{AP}}(\vec\lambda^{\text{align}})} \cdot \nn
    && \Big(\partial_{\lambda_i} h^{\text{AP}}(\vec\lambda^{\text{align}}) \Big| h^s (\vec\lambda^{\text{tr}}) - h^{\text{AP}}(\vec\lambda^{\text{align}}) \Big)\,.
\end{eqnarray}
In this work, we focus on three key parameters: the chirp mass $\mathcal M_c$, the symmetric mass ratio $\eta$, and the luminosity distance $d_L$. The extrinsic parameters $t_c$ and $\phi_c$ are aligned \textit{a priori} in the FCV implementation and the mismatch computation. We demonstrate in Appendix~\ref{appx:wa}, the aligned waveform generally achieves a lower mismatch than nonaligned waveforms, thereby extending the region of validity for the FCV method.

\section{Systematic Bias Induced by Omitting Eccentricity} \label{sec:results}

\subsection{B-DECIGO} \label{sec:fcv-bdecigo}

\begin{figure*}[htbp]
    \centering
    \includegraphics[width=0.9\linewidth]{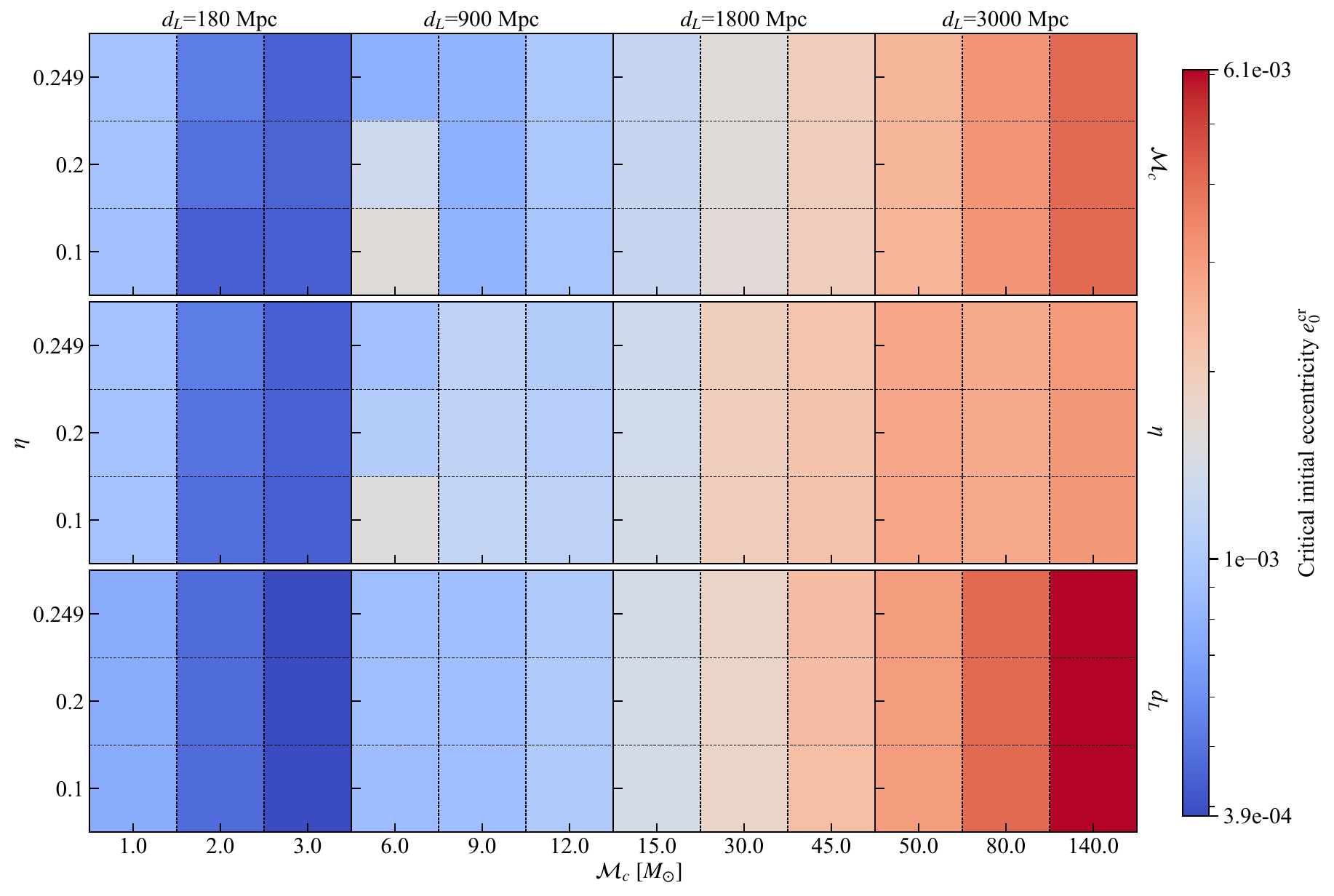}
    \caption{The critical initial eccentricity $e_{0}^{\text{cr}}$ for $\mathcal M_c$, $\eta$ and $d_L$ based on a B-DECIGO catalog spanning a range of these parameters. From top to bottom, the panels correspond to $e_{0}^{\text{cr}}$ for $\mathcal{M}_c$, $\eta$, and $d_L$, respectively.}
    \label{decigo-r1}
\end{figure*}
\begin{table*}[t]
\centering
\caption{Chirp mass $\mathcal M_c$, symmetric mass ratio $\eta$, luminosity distance $d_L$, SNR, starting frequency $f_{\text{start}}$, in-band duration $T_{\text{in-band}}$, number of in-band cycles and critical eccentricities $e_0^{\text{cr}}$ for selected events in the GW catalog based on B-DECIGO.}
\label{dcg_event}
\begin{tabular}{>{\centering\arraybackslash}p{65pt}
                >{\centering\arraybackslash}p{35pt}
                >{\centering\arraybackslash}p{20pt}
                >{\centering\arraybackslash}p{40pt}
                >{\centering\arraybackslash}p{25pt}
                >{\centering\arraybackslash}p{40pt}
                >{\centering\arraybackslash}p{45pt}
                >{\centering\arraybackslash}p{45pt}
                >{\centering\arraybackslash}p{45pt}
                >{\centering\arraybackslash}p{45pt}
                >{\centering\arraybackslash}p{45pt}}
\hline
Event & $\mathcal M_c\ [M_\odot]$ & $\eta$ & $d_L$ [Mpc] & SNR & $f_{\text{start}}$ [Hz] & $T_{\text{in-band}}$[yr] & GW cycles & $e_{0}^{\text{cr}}(\mathcal M_c)$ & $e_{0}^{\text{cr}}(\eta)$ & $e_{0}^{\text{cr}}(d_L)$ \\
\hline
GW231123-like & 153.60 & 0.248 & 2700 & 85.12  & 0.10 & 0.00216 & 10903 & 4.070e-03 & 3.008e-03 & 6.240e-03 \\
GW190521-like & 101.15 & 0.232 & 3310 & 49.02  & 0.10 & 0.00433 & 21873 & 3.649e-03 & 2.925e-03 & 5.030e-03 \\
GW150914-like & 30.84  & 0.249 & 470  & 128.29 & 0.10 & 0.03138 & 158389 & 1.030e-03 & 1.216e-03 & 1.147e-03 \\
GW190814-like & 6.43   & 0.090 & 230  & 72.29  & 0.10 & 0.42848 & 2162493 & 5.469e-04 & 5.622e-04 & 4.949e-04 \\
GW200115-like & 2.57   & 0.158 & 290  & 25.92  & 0.13 & 1.00000 & 6513054 & 6.368e-04 & 6.406e-04 & 4.933e-04 \\
GW170817-like & 1.20   & 0.249 & 40   & 82.65  & 0.21 & 1.00000 & 10488891 & 3.206e-04 & 3.598e-04 & 4.137e-04 \\
\hline
\end{tabular}
\end{table*}
\begin{figure*}[htbp]
    \centering
    \includegraphics[width=0.9\linewidth]{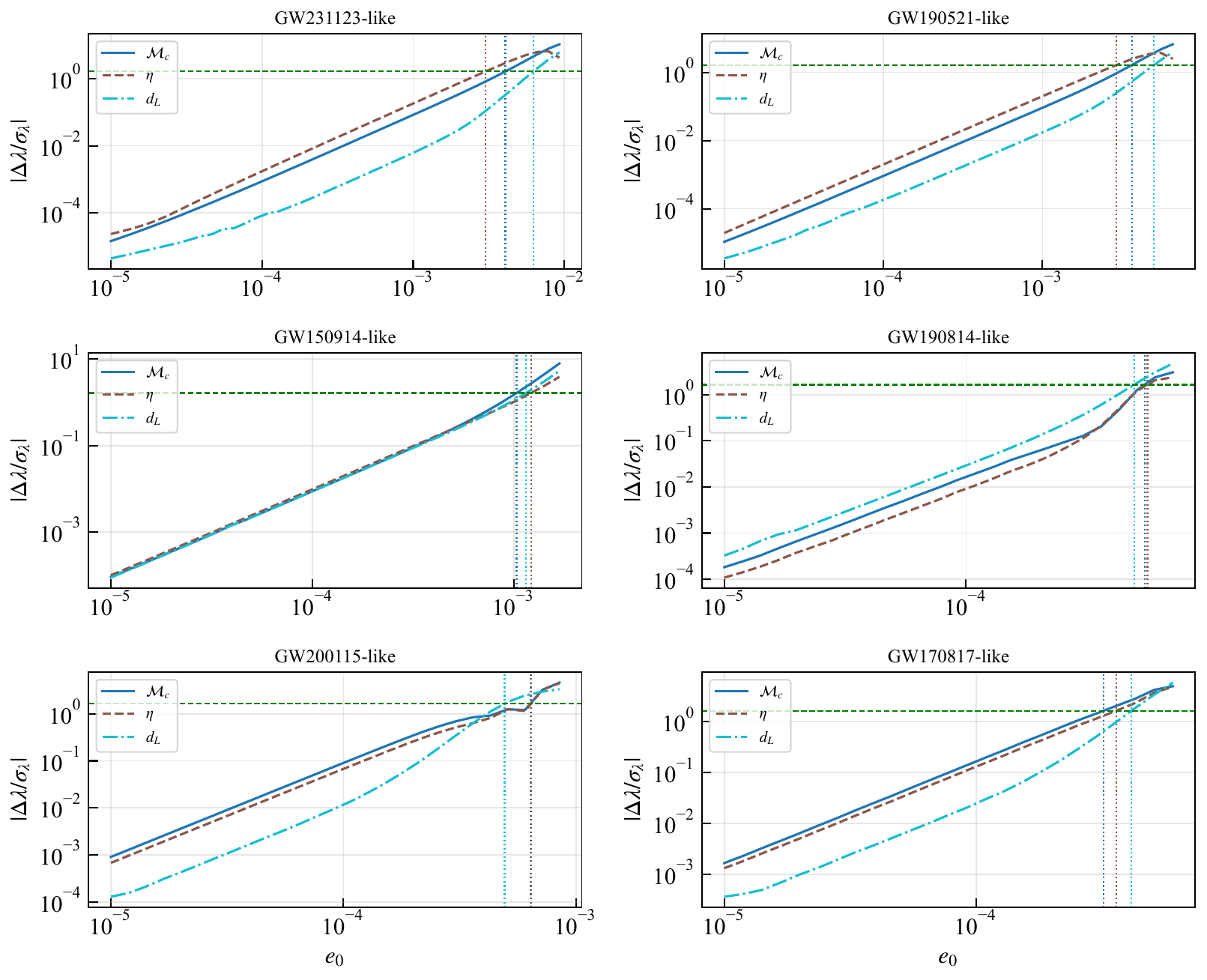}
    \caption{Normalized FCV systematic biases $|\Delta \vec\lambda^{\text{sys}} / \sigma|$ for $\mathcal M_c$, $\eta$ and $d_L$ as functions of initial eccentricity $e_{0}$ for the six typical GW events. The vertical dotted line indicates the location of the critical eccentricity $e_{0}^{\text{cr}}$ on the axes. The green dashed line denotes the threshold value of 1.645.}
    \label{decigo1}
\end{figure*}

In the deci-Hertz band, analyzing eccentric GW signals with quasicircular templates can induce non-negligible systematic errors. Moreover, compared with analyzes based on the LVK band, even smaller eccentricities can lead to significant biases. 
We construct catalogs of events spanning a broad region of parameter space, in which the detector-frame chirp mass $\mathcal M_c$ ($M_\odot$), symmetric mass ratio $\eta$, and luminosity distance $d_L$ (Mpc) are either sampled over representative intervals or chosen from selected events in GWTC \cite{LIGOScientific:2025hdt, LIGOScientific:2025slb, LIGOScientific:2025snk, KAGRA:2023pio, LIGOScientific:2019lzm}.
Our catalogs include all three types of CBCs and span a typical mass range from $\mathcal O(1)$ to $\mathcal O(100)\,M_\odot$.
Eccentric GW signals are generated and, after applying waveform alignment, we compute values of the FCV biases, normalizing them with respect to the $1\sigma$ statistical errors derived from the covariance matrix.
To suppress the influence of angular parameters on the FCV biases of the chirp mass, symmetric mass ratio, and luminosity distance, we randomly sample 1000 sets of angular parameters $(\iota, \theta, \phi, \psi, \beta)$ within their allowed ranges, with $(\theta, \phi)$ drawn uniformly over the celestial sphere. The results are then averaged over these 1000 realizations. 
Without loss of generality, the injected coalescence time and phase are fixed to $t_c = 0$ and $\phi_c = 0$.
The reference frequency is set to $f_0 = 0.1$ Hz.

Figure~\ref{decigo-r1} presents the dependence of the critical eccentricities $e_{0}^{\text{cr}}$ for $\mathcal M_c$, $\eta$ and $d_L$, based on a catalog spanning a range of these parameters. The values of $e_{0}^{\text{cr}}$ are computed by interpolating the curves of $|\Delta \vec\lambda^{\text{sys}} / \sigma|$ versus $e_{0}$ and numerically approximating their intersection with $|\Delta \vec\lambda^{\text{sys}} / \sigma| = 1.645$. There is a clear trend that $e_{0}^{\text{cr}}$ increases with increasing $\mathcal M_c$. Larger luminosity distances $d_L$ are associated with higher values of $e_{0}^{\text{cr}}$. As we demonstrate in Fig.~\ref{bd_snr}, the SNRs of events with different $d_L$ remain within the same order of magnitude, indicating that this trend is not driven by large variations in SNR. Overall, the critical eccentricities lie in the range $\mathcal{O}(10^{-4})$–$\mathcal{O}(10^{-3})$.

Figure~\ref{decigo1} shows the normalized FCV biases, $|\Delta \vec\lambda^{\text{sys}} / \sigma|$, for $\mathcal M_c$, $\eta$, and $d_L$ as functions of the initial eccentricities $e_{0}$, based on six selected events from GWTC. Table~\ref{dcg_event} also lists the corresponding critical eccentricities $e_{0}^{\text{cr}}$, defined as the initial eccentricity at which the normalized FCV bias first exceeds $1.645$.
The in-band duration $T_{\text{in-band}}$ and the number of in-band cycles are reported for the quasicircular case, since the eccentricities considered here only slightly change the duration and accumulated cycle count.
One can observe an approximately monotonic increase of the normalized FCV biases as the initial eccentricity $e_{0}$ grows, with the value of biases reaching the boundary of the 90\% CL when $e_{0}$ lies in the range $\mathcal{O}(10^{-4})$ to $\mathcal{O}(10^{-3})$, depending on the specific injected parameter values.

\subsection{LISA} \label{sec:fcv-lisa}

\begin{figure*}[htbp]
    \centering
    \includegraphics[width=0.9\linewidth]{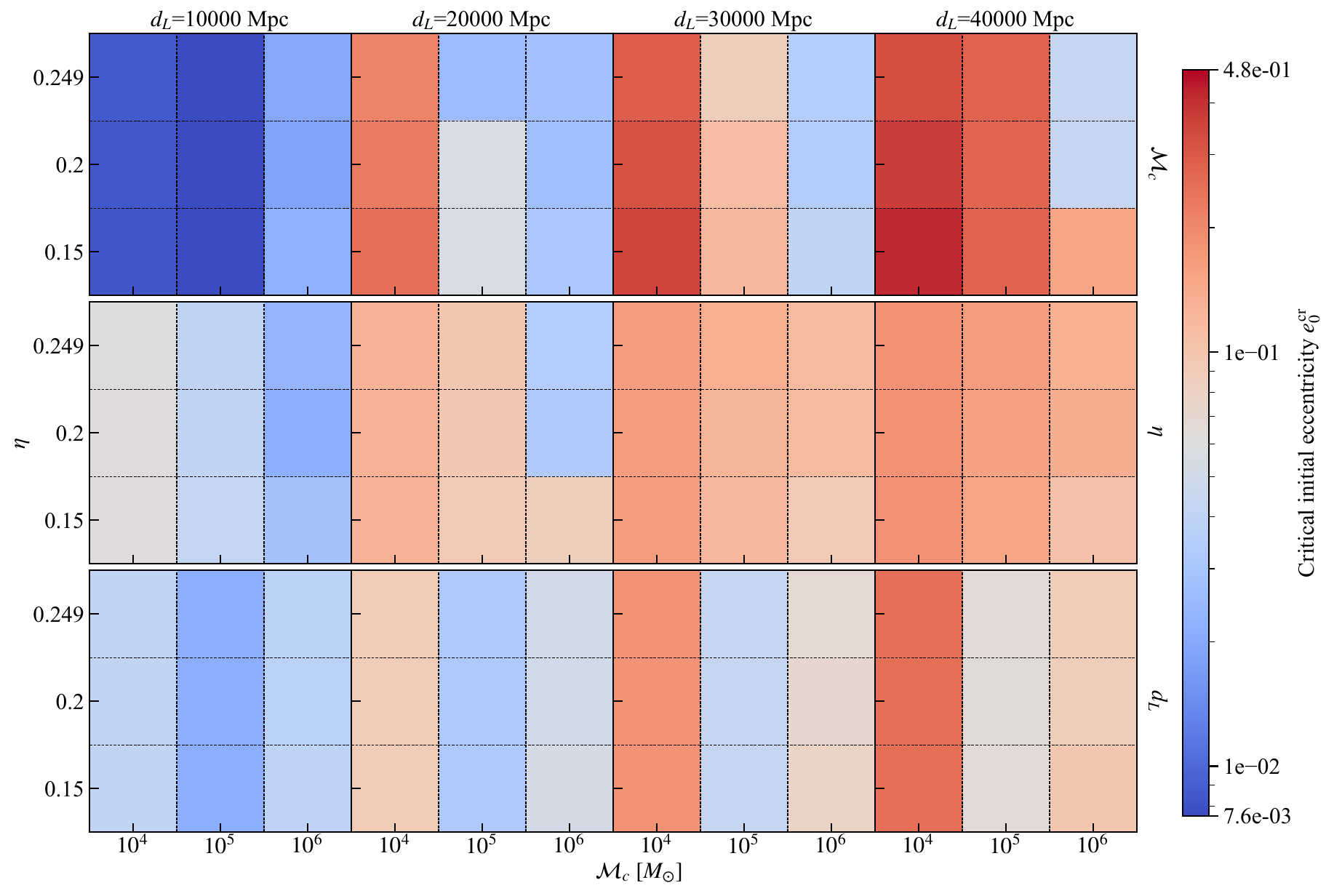}
    \caption{The critical initial eccentricity $e_{0}^{\text{cr}}$ for $\mathcal M_c$, $\eta$ and $d_L$,  based on a LISA catalog spanning a range of these parameters. From top to bottom, the panels correspond to $e_{0}^{\text{cr}}$ for $\mathcal{M}_c$, $\eta$, and $d_L$, respectively.}
    \label{lisa-r1}
\end{figure*}
\begin{table*}[t]
\centering
\caption{Chirp mass $\mathcal M_c$, symmetric mass ratio $\eta$, luminosity distance $d_L$, SNR, starting frequency $f_{\text{start}}$, in-band duration $T_{\text{in-band}}$, number of in-band cycles and critical eccentricities $e_0^{\text{cr}}$ for selected events in the GW catalog based on LISA. Note that for LISA6, $e_{0}^{\text{cr}}(\mathcal{M}_c)$ is not reached within the explored eccentricity range, owing to comparatively large statistical uncertainties arising from its limited SNR.}
\label{lisat_event}
\begin{tabular}{>{\centering\arraybackslash}p{25pt}
                >{\centering\arraybackslash}p{40pt}
                >{\centering\arraybackslash}p{30pt}
                >{\centering\arraybackslash}p{40pt}
                >{\centering\arraybackslash}p{30pt}
                >{\centering\arraybackslash}p{40pt}
                >{\centering\arraybackslash}p{45pt}
                >{\centering\arraybackslash}p{45pt}
                >{\centering\arraybackslash}p{45pt}
                >{\centering\arraybackslash}p{45pt}
                >{\centering\arraybackslash}p{45pt}}
\hline
Event & $\mathcal M_c\ [M_\odot]$ & $\eta$ & $d_L$ [Mpc] & SNR & $f_{\text{start}}$ [Hz] & $T_{\text{in-band}}$[yr] & GW cycles & $e_{0}^{\text{cr}}(\mathcal M_c)$ & $e_{0}^{\text{cr}}(\eta)$ & $e_{0}^{\text{cr}}(d_L)$ \\
\hline
LISA1 & 3.896e5 & 0.215 & 57025.65 & 291.33 & 0.00010 & 0.45778 & 2307 & 2.385e-01 & 1.500e-01 & 1.138e-01 \\
LISA2 & 5.774e5 & 0.204 & 29289.09 & 409.04 & 0.00010 & 0.23771 & 1196 & 3.801e-02 & 1.083e-01 & 5.931e-02 \\
LISA3 & 1.355e6 & 0.233 & 28840.34 & 178.93 & 0.00010 & 0.05732 & 286 & 3.670e-02 & 4.928e-02 & 7.220e-02 \\
LISA4 & 3.272e6 & 0.239 & 54890.07 & 65.64  & 0.00010 & 0.01307 & 63 & 1.680e-01 & 1.540e-01 & 2.715e-01 \\
LISA5 & 1.623e4 & 0.246 & 19340.53 & 101.92 & 0.00030 & 5.00000 & 75094 & 2.170e-01 & 1.231e-01 & 5.332e-02 \\
LISA6 & 2.447e3 & 0.193 & 28921.43 & 14.38  & 0.00097 & 5.00000 & 244855 & -- & 2.432e-01 & 1.615e-01 \\
\hline
\end{tabular}
\end{table*}
\begin{figure*}[htbp]
    \centering
    \includegraphics[width=0.9\linewidth]{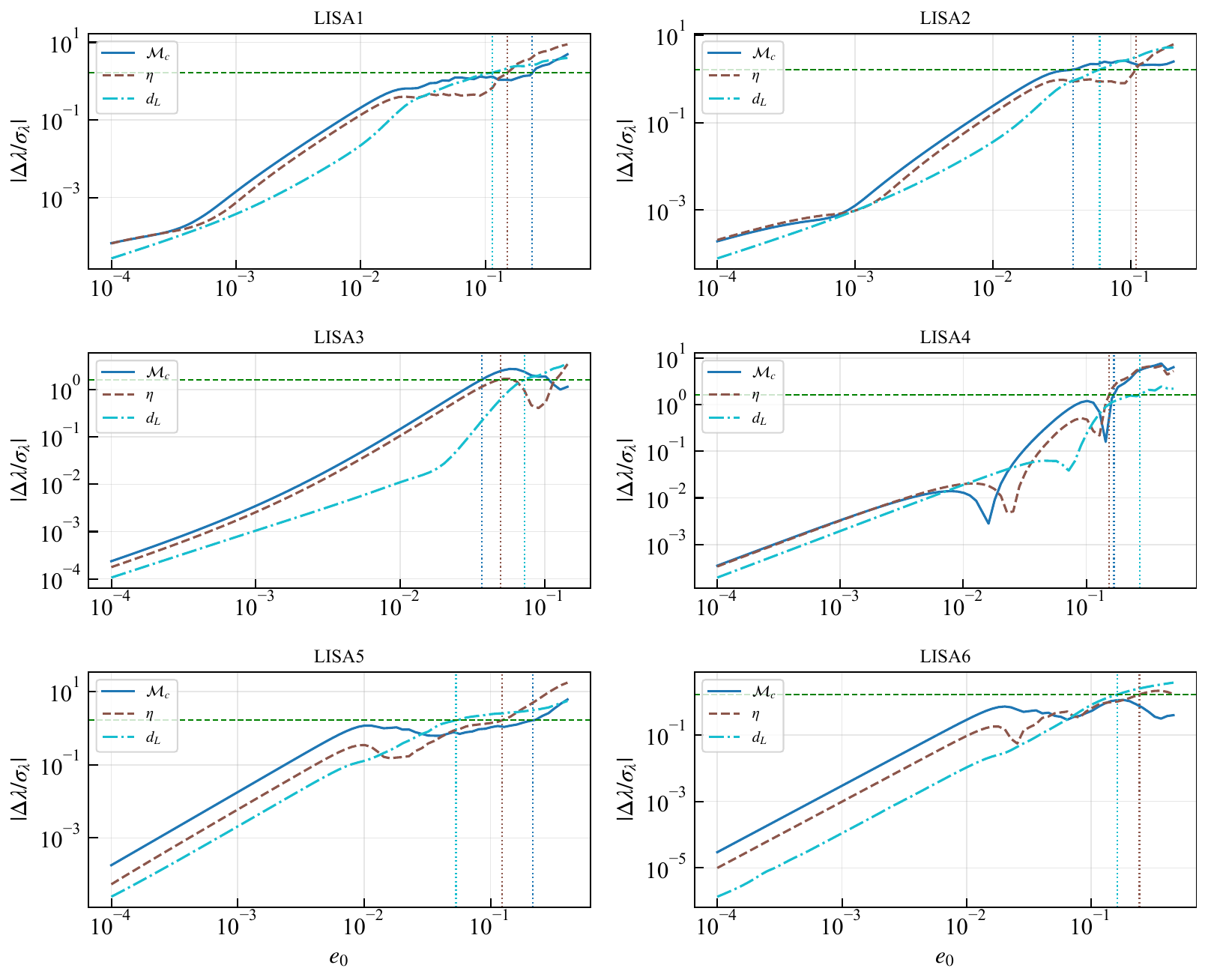}
    \caption{Normalized FCV systematic biases $|\Delta \vec\lambda^{\text{sys}} / \sigma|$ for $\mathcal M_c$, $\eta$ and $d_L$ as functions of initial eccentricity $e_{0}$ for the six typical GW events. The vertical dotted line indicates the location of the critical eccentricity $e_{0}^{\text{cr}}$ on the axes. The green dashed line denotes the threshold value of 1.645.}
    \label{lisa1}
\end{figure*}

Based on the LISA catalog for massive black hole binaries (MBHBs) proposed in \cite{Klein:2015hvg}, we construct two catalogs of events: one spanning a broad range of parameters, including detector-frame chirp mass $\mathcal M_c$ ($M_\odot$), symmetric mass ratio $\eta$, and luminosity distance $d_L$ (Mpc), and another consisting of six selected MBHBs that distinguish between light and heavy seeding mechanisms, and a detection threshold of $\text{SNR} = 8$ is applied.

Following the same procedure as for B-DECIGO, we generate injected eccentric GW signals, align the waveform, compute the FCV biases, and normalize them by the 1-$\sigma$ statistical errors. The results are averaged over 1000 realizations of angular parameters, which are generated in the same manner as for the B-DECIGO case, while $t_c = 0$ and $\phi_c = 0$ are fixed for the injections. The reference frequency is set to $f_0 = 0.0001$ Hz.

We first construct a catalog in which $\mathcal{M}_c$, $\eta$, and $d_L$ span representative ranges. The resulting variation of $e_{0}^{\text{cr}}$ is shown in Fig.~\ref{lisa-r1}, with the corresponding SNRs presented in Fig.~\ref{ls_snr}. We find that $e_{0}^{\text{cr}}$ increases with increasing $d_L$.
Several mass-related factors influence the value of $e_{0}^{\text{cr}}$. For example, at a fixed luminosity distance, more massive binaries would generally yield higher SNRs. However, once the cutoff frequency $f_{\text{ISCO}}$ is taken into account, this trend does not necessarily hold, as higher-mass binaries have lower $f_{\text{ISCO}}$. Moreover, more massive binaries spend fewer cycles in band during the inspiral stage. As a result, the dependence of $e_{0}^{\text{cr}}$ on $\mathcal{M}_c$ is not monotonic.
Overall, $e_{0}^{\text{cr}}$ lies in the range $\mathcal{O}(10^{-2})$–$\mathcal{O}(10^{-1})$ at $f_0 = 0.0001$ Hz.

For the catalog of selected events, the parameters, including detector-frame chirp mass $\mathcal M_c$, symmetric mass ratio $\eta$, and luminosity distance $d_L$, are summarized in Table~\ref{lisat_event}.
In this catalog, LISA1–LISA4 correspond to heavy-seed scenario, spanning a mass range from $\mathcal O(10^5)$ to $\mathcal O(10^6)\,M_\odot$, and LISA5–LISA6 correspond to light seeds, with masses spanning from $\mathcal O(10^3)$ to $\mathcal O(10^4)\,M_\odot$. These events reflect the typical mass ranges expected for the two seeding scenarios.
It is noteworthy that LISA3 and LISA4, despite having larger chirp masses, exhibit smaller SNRs due to their much lower ISCO frequencies: $f_{\text{ISCO}} = 0.001352$ Hz and $0.0005687$ Hz for LISA3 and LISA4, respectively, compared with $0.004488$ Hz and $0.002932$ Hz for LISA1 and LISA2.
Figure~\ref{lisa1} presents the normalized FCV biases $|\Delta \vec\lambda^{\text{sys}} / \sigma|$ for $\mathcal M_c$, $\eta$, and $d_L$ as functions of the initial eccentricities $e_{0}$, based on the six selected MBHB events. Table~\ref{lisat_event} summarizes the corresponding critical eccentricities $e_{0}^{\text{cr}}$. The in-band duration $T_{\text{in-band}}$ and the number of in-band cycles in Table~\ref{lisat_event} are likewise reported for the quasicircular case, for the same reason as in the B-DECIGO case.
An overall increasing trend is observed for the normalized FCV biases as the initial eccentricity $e_{0}$ grows. For larger values of $e_{0}$, oscillatory behavior appears in the estimated parameter shifts $\Delta \vec\lambda^{\text{sys}}$. This arises because the linearized approximation of the exponential term for the phase differences, $h^s_{\text{ecc}} - h^{\text{AP}}_{\text{qc}} \simeq i\mathcal A (\Psi_{\text{ecc}} - \Psi_{\text{qc}})$, where amplitude corrections are neglected as subdominant to phase corrections, ceases to be valid near the aligned parameter set $\vec\lambda^{\text{align}}$.
The biases reach the boundary of the 90\% CL when $e_{0}$ lies in the range $\mathcal{O}(10^{-2})$–$\mathcal{O}(10^{-1})$, depending on the specific injection parameters.
Overall, systematic biases are less pronounced in LISA than in B-DECIGO, primarily because LISA’s lower sensitive frequency band results in fewer in-band GW cycles.

\subsection{Validation of FCV results using Bayesian inference} \label{validation}

\begin{figure*}
    \centering
    \includegraphics[width=0.8\linewidth]{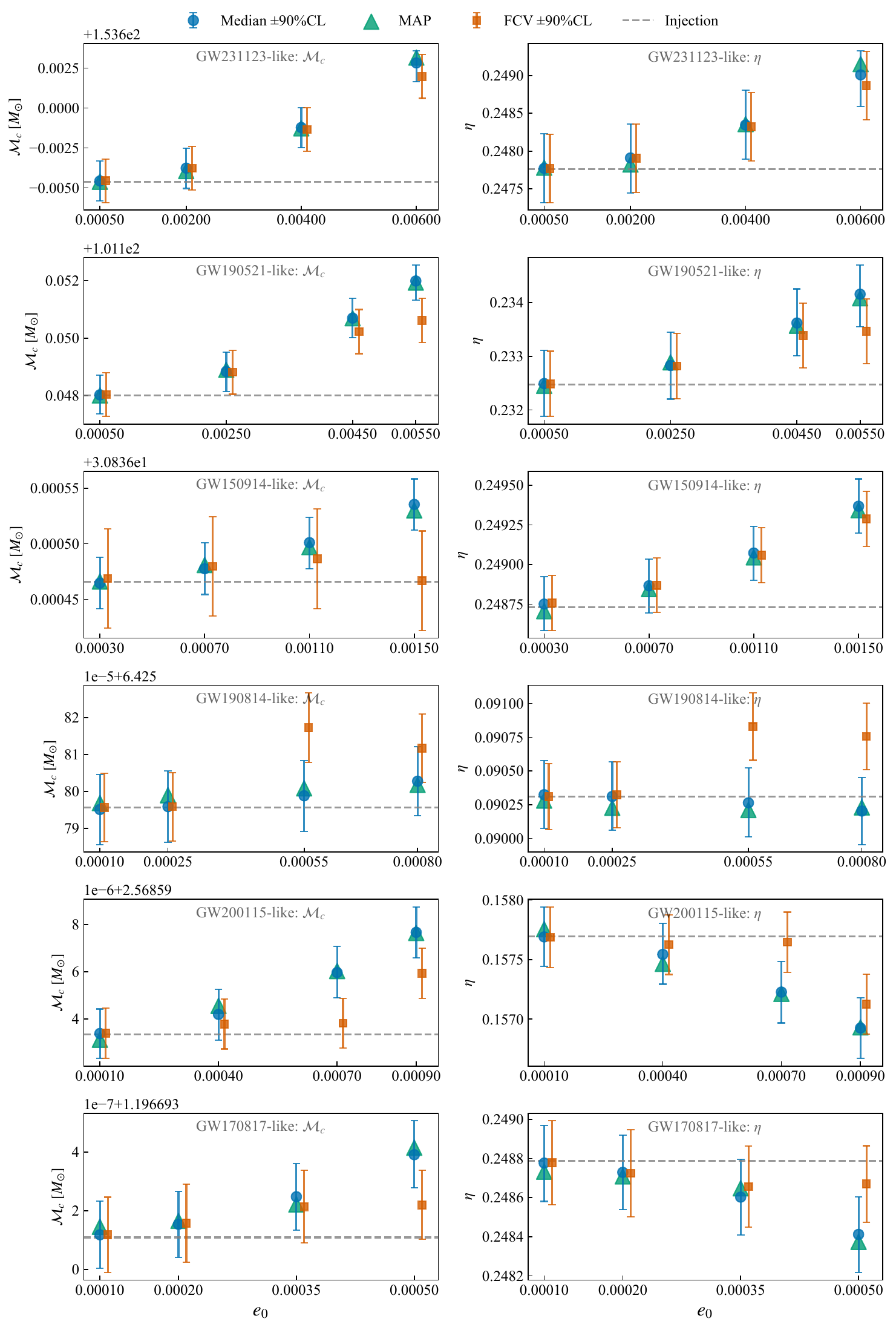}
    \caption{Comparison between the FCV, FIM and Bayesian results for systematic and statistical errors of B-DECIGO events. The FIM and FCV results are slightly shifted to the right for visual clarity. Bayesian estimates are shown in blue (MAP) and green (posterior median), while the FCV and FIM forecasts are indicated by orange lines and markers. The dotted line denotes the injected parameter value.}
    \label{bayes1}
\end{figure*}

\begin{figure*}
    \centering
    \includegraphics[width=0.8\linewidth]{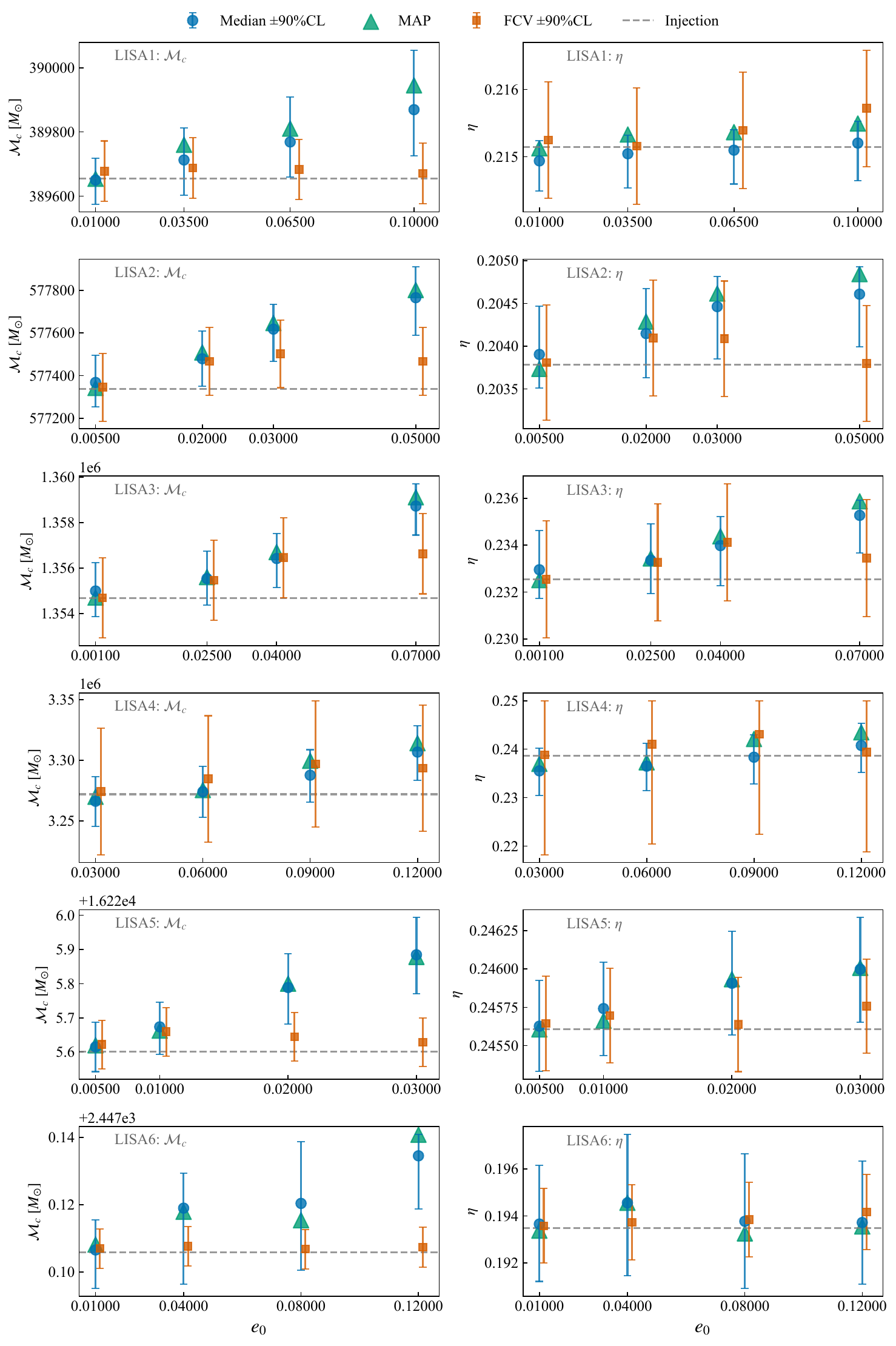}
    \caption{Comparison between the FCV, FIM and Bayesian results for systematic and statistical errors of LISA events. The FIM and FCV results are slightly shifted to the right for visual clarity. Bayesian estimates are shown in blue (MAP) and green (posterior median), while the FCV and FIM forecasts are indicated by orange lines and markers. The dotted line denotes the injected parameter value.}
    \label{bayes2}
\end{figure*}

To assess the validity of the FCV and FIM predictions, we perform zero-noise injection-recovery studies and Bayesian analyses.
The angular parameters are fixed to $(\iota, \theta, \phi, \psi, \beta) = (\pi/4, \pi/4, \pi/4, \pi/4, \pi/4)$, with the injected values of $t_c$ and $\phi_c$ fixed to zero. All inferences are carried out on a 32-core server using frequency-domain waveforms. 
For each source, we consider four increasing values of the initial eccentricity $e_0$.
We inject eccentric signals using EccentricFD and adopt the same approximant with setting $e_0 = 0$ as the recovery template, and infer all nine quasicircular parameters ${\mathcal M_c, \eta, d_L, \iota, \theta, \phi, \psi, t_c, \phi_c}$ for the selected B-DECIGO and LISA events. Priors are taken to be uniform within $\pm 10\sigma$ of the injected values, subject to physical constraints (e.g., $0 < \eta \leq 0.25$), where the $\sigma$ values are estimated using the FIM. 
The resulting posteriors are shown in Figs.~\ref{bayes1} and \ref{bayes2}.
For each parameter, we report the posterior median and the 90\% CL, while the MAP estimate is identified as the location of the maximum posterior density in the full parameter space.
For visual clarity, the FIM and FCV forecast results are slightly shifted to the right. For FCV $\pm$90\% CL of $\eta$ in the event LISA4, the upper error bar is truncated at the physical limit $\eta = 0.25$.

As $e_0$ increases, systematic deviations from the injected values become progressively more pronounced.
For small $e_0$, the FCV predictions are in good agreement with the Bayesian results, in terms of both systematic biases and statistical uncertainties. Minor discrepancies in the FIM-predicted statistical uncertainties are observed only for the GW150914-like, LISA4, and LISA6 cases.
At larger $e_0$, the bias estimates obtained from the two methods begin to show differences.
From Figs.~\ref{bayes1} and \ref{bayes2}, 10 out of the 12 selected systems exhibit differences exceeding the Bayesian 90\% CL bound when $e_0$ is above certain values: $e_0 = 0.0055$ for the GW190521-like system, $e_0 = 0.0015$ for the GW150914-like system, $e_0 = 0.00055$ for the GW190814-like system, $e_0 = 0.0007$ for the GW200115-like system, $e_0 = 0.0005$ for the GW170817-like system, $e_0 = 0.1$ for the LISA1 system, $e_0 = 0.05$ for the LISA2 system, $e_0 = 0.07$ for the LISA3 system, $e_0 = 0.02$ for the LISA5 system, and $e_0 = 0.12$ for the LISA6 system.
These results indicate that the FCV framework provides accurate bias estimates provided that $e_0$ remains below the values identified above. At larger initial eccentricities, the accuracy of FCV predictions may degrade, driven by the increasing mismatch between eccentric and quasicircular waveforms (see Figs.~\ref{mmnabd} and \ref{mmnals}), which signals a breakdown of the linear approximation that the FCV method should fulfill.
It is also worth noting that for the GW190521-like, GW190814-like, and GW170817-like systems, as well as the LISA2 and LISA3 systems, the systematic biases exceed the FIM-predicted 90\% CL bounds at the corresponding $e_0$ values identified above. This further supports the robustness of the FCV-predicted critical eccentricity $e_0^{\text{cr}}$ for these events.

The comparisons between FCV and Bayesian analysis presented here are obtained for a specific set of angular parameters and may therefore not be generalizable to other configurations. Due to the high computational cost, we defer a more comprehensive validation of FCV across a broader range of angular parameters to future work.

Overall, the FCV and FIM predictions show good agreement with Bayesian results over most of the explored $e_0$ range. However, at larger $e_0$, the systematic biases inferred from FCV and Bayesian analyses begin to differ, indicating that FCV-based forecasts may lose accuracy in this regime, and that Bayesian inference should be adopted instead.
Within the range of initial eccentricities considered in this work, the FCV method nonetheless provides a reliable and consistent interpretation for the significance of systematic errors.

\begin{figure*}[htbp]
  \centering
  \begin{minipage}[t]{0.48\linewidth}
      \centering
      \includegraphics[width=\linewidth]{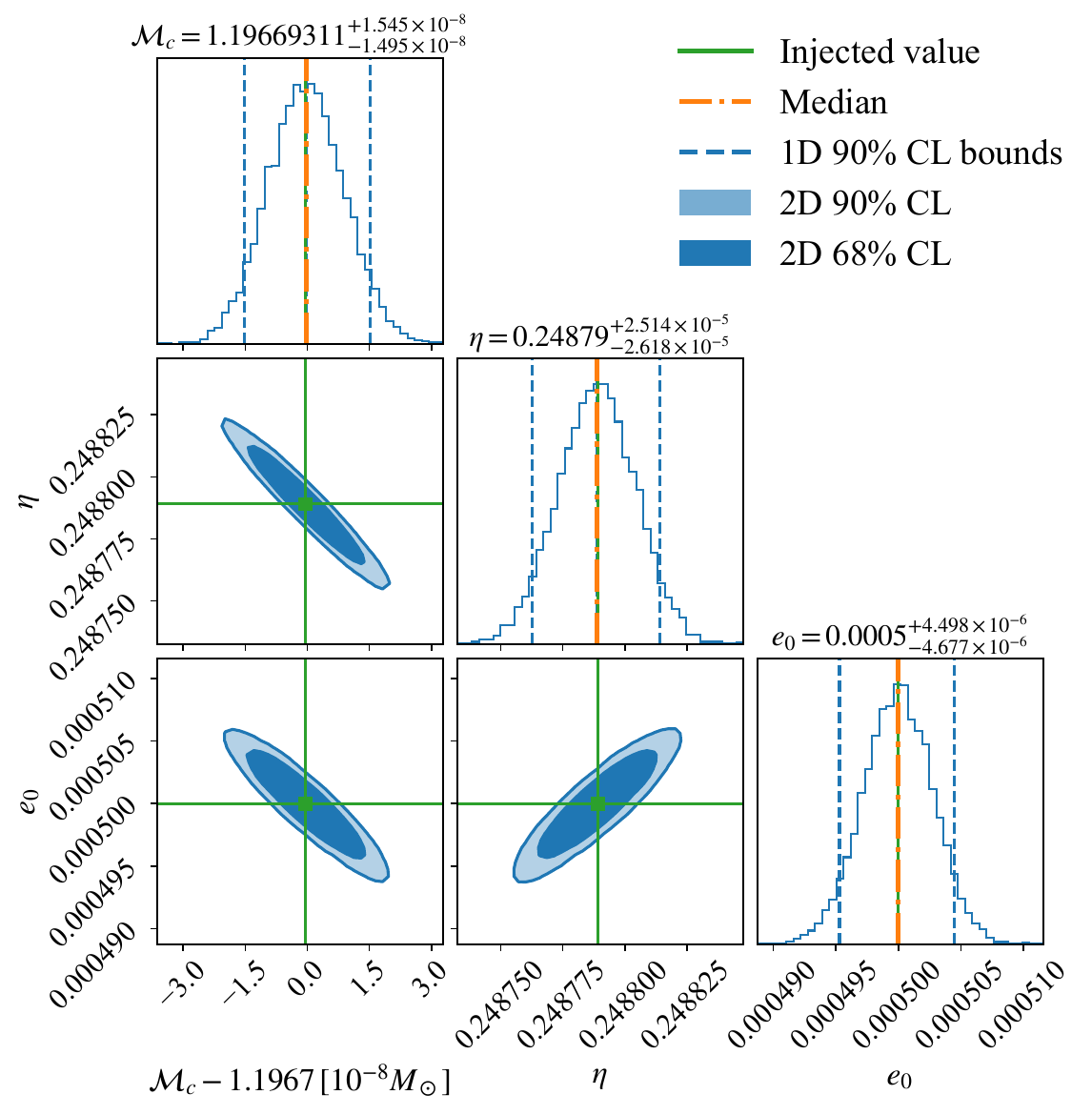}
  \end{minipage}
  \hfill
  \begin{minipage}[t]{0.48\linewidth}
      \centering
      \includegraphics[width=\linewidth]{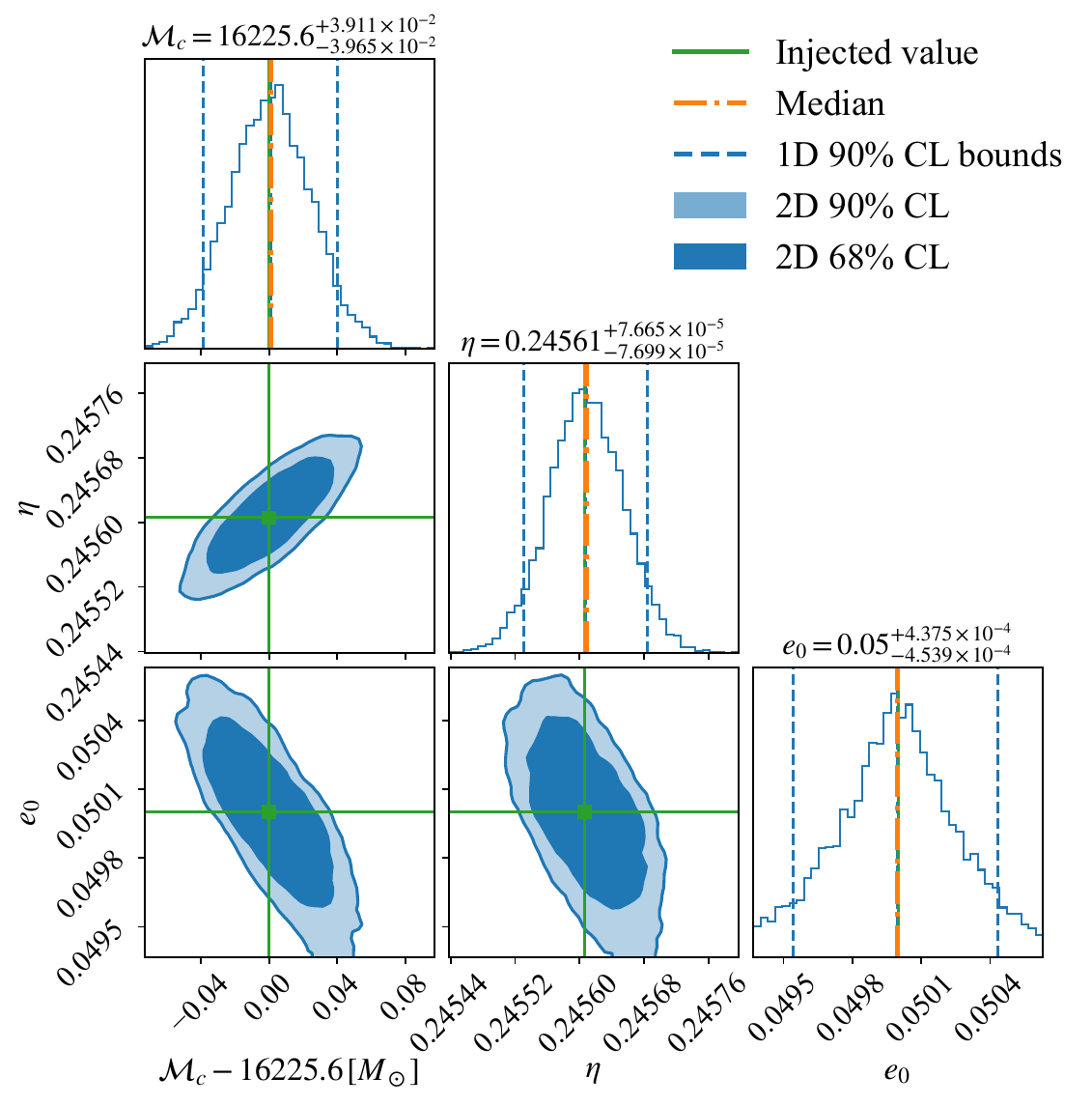}
  \end{minipage}
  \caption{The posteriors of $\mathcal M_c$, $\eta$ and eccentricity $e_0$. Left: the GW170817-like injection observed with B-DECIGO; Right: the LISA5 injection observed with LISA. The green solid, orange dash-dotted, and blue dashed vertical lines in the one-dimensional posterior distribution indicate the injected, median, and bounds of the 90\% credible interval, respectively. The darker and lighter shaded regions denote the 68\% and 90\% credible intervals of the two-dimensional posterior distribution, respectively.}
  \label{fig:ecc-recov}
\end{figure*}

Moreover, as a validation of the methodology, we perform Bayesian analyses in which eccentric signals are injected and recovered using the same EccentricFD template, with eccentricity treated as a free parameter. In these runs, the total number of inferred parameters is 11: the eccentricity-related parameters $e_0$ and $\beta$, together with the nine parameters used in the main analysis. We consider two representative systems, a GW170817-like event for B-DECIGO and the LISA5 event for LISA. The injected angular parameters, as well as $t_c$ and $\phi_c$, are chosen with the same configurations described above. The injected eccentricity is chosen to be close to the corresponding critical eccentricity $e_0^{\text{cr}}$.
In both cases, the injected parameters are recovered without biases. As representative examples, we present the posteriors of $\mathcal{M}_c$, $\eta$, and the eccentricity $e_0$ for these two events in Fig.~\ref{fig:ecc-recov}. For the GW170817-like injection, the lower and upper uncertainties of the one-dimensional 90\% CL for $e_0$, relative to the median, are $4.677\times 10^{-6}$ and $4.498\times 10^{-6}$, respectively. For the LISA5 injection, the corresponding values are $4.539\times 10^{-4}$ and $4.375\times 10^{-4}$. These uncertainties are much smaller than the injected eccentricities, $e_0=5\times 10^{-4}$ and $e_0=5\times 10^{-2}$, respectively. 
These unbiased injection--recovery results and the tight constraints on eccentricity provide an additional consistency check of the methodology used to determine $e_0^{\text{cr}}$.


\section{Conclusion and Remarks} \label{sec:conclusion}

In this work, we systematically investigated the impact of eccentricity on parameter estimation for compact binary coalescence (CBC) mergers based on two spaceborne detectors: the decihertz B-DECIGO and the millihertz LISA. Our analysis accounted for the additional harmonics induced by eccentricity, the corresponding detector responses, and the orbital motion of the spaceborne detectors over their year-long duty cycles. By employing both the FCV method and Bayesian inference, we quantified the systematic biases arising from mismatched waveform templates due to the omission of eccentricity.

Our results show that even very small initial eccentricities can produce non-negligible systematic biases in key parameters such as the chirp mass $\mathcal{M}_c$, symmetric mass ratio $\eta$, and luminosity distance $d_L$. We introduced the critical eccentricity, $e_{0}^{\rm cr}$, to mark the threshold at which systematic errors become comparable to the statistical errors, as measured by normalized FCV biases. 
In the B-DECIGO band, significant biases appear at initial eccentricities $e_{0} \sim \mathcal{O}(10^{-4})$--$\mathcal{O}(10^{-3})$, 
whereas in LISA, where the in-band frequency is lower and fewer waveform cycles are present, systematic effects generally emerge at larger eccentricities $e_{0} \sim \mathcal{O}(10^{-2})$--$\mathcal{O}(10^{-1})$.

Using a set of selected events with fixed angular parameters, comparisons between FCV predictions and Bayesian inference demonstrate that the FCV method provides accurate and consistent estimates of systematic biases for eccentricities up to the values of $e_0$ identified in Sec.~\ref{validation}. Noticeable deviations arise primarily at higher eccentricities; however, the FCV method still captures the relative significance of the biases in this regime.
In this work, we focus on employing the FCV framework to forecast the critical eccentricity $e_0^{\text{cr}}$, and we leave a detailed investigation of the inconsistencies between FCV and Bayesian results to future studies. 
Moreover, we show that prealigning waveforms by optimizing the extrinsic parameters ($t_c$ and $\phi_c$) reduces waveform mismatches and extends the validity range of the FCV approximation.

We compare our results with previous studies.
It should be noted that, for B-DECIGO, we adopt a one-year observation period with $f_{\text{min}} = 0.1$ Hz, while for LISA we adopt a five-year observation period with $f_{\text{min}} = 10^{-4}$ Hz.
For B-DECIGO, \cite{GilChoi:2022waq} reported that eccentricities $e_{0} \sim \mathcal{O}(10^{-4})$ can already produce systematic errors comparable to statistical uncertainties. In this work, we improve the modeling accuracy by incorporating the time-dependent detector pattern function and deriving the detector response separately for each harmonic in the waveform.
For LISA, \cite{Garg:2024oeu} found that $e_{0} \sim \mathcal{O}(10^{-2})$--$\mathcal{O}(10^{-1.5})$ leads to significant parameter biases, while \cite{Garg:2024qxq} showed that $e_{0} \gtrsim \mathcal{O}(10^{-2})$ can induce apparent deviations from general relativity. In this work, we consider an eccentric waveform approximant EccentricFD, implementing eccentricity-induced harmonic structure. Despite differences in waveform modeling and detector response, our results are in good agreement with previous findings at the order-of-magnitude level.

In this work, we adopt the frequency-domain, inspiral-only, nonspinning EccentricFD waveform. Consequently, spin effects are neglected, and higher-order multipole modes, which can be important for asymmetric mass ratios and nonzero inclination angles, are not included.
We also do not model the merger-ringdown stage, for which eccentricity has not yet been consistently incorporated into the waveform of this stage.
Fully consistent frequency-domain inspiral-merger-ringdown (IMR) models that simultaneously incorporate eccentricity, spin precession, and higher-order modes are still under active development.
A more comprehensive analysis, employing a physically complete, computationally efficient waveform model is left for future work.

Overall, our results emphasize the necessity of properly accounting for eccentricity in waveform models, especially for future spaceborne detectors. The FCV method, particularly when combined with waveform alignment, provides a computationally efficient and robust tool for predicting systematic biases induced by residual eccentricity. This is crucial for the accurate recovery of source parameters, including masses, distances, and orbital characteristics, in future deci-Hertz and milli-Hertz gravitational-wave observations. Properly incorporating eccentricity will improve parameter estimation fidelity, reduce the risk of misinterpreting astrophysical information about GW sources, and enhance the scientific potential of upcoming space-based GW missions.

\section*{Acknowledgments}
This work is supported by the National Natural Science Foundation of China Grant No. 12575063, and in part by ``the Special Funds for the Double First-Class Development of Wuhan University'' under the reference No. 2025-1302-010. Part of the numerical calculations in this paper have been done on the supercomputing system in the Supercomputing Center of Wuhan University.

\section*{Data availability}

The data that support the findings of this article are not publicly available. The data are available from the authors upon reasonable request.

\appendix

\begin{figure*}[htbp]
    \centering
    \includegraphics[width=0.85\linewidth]{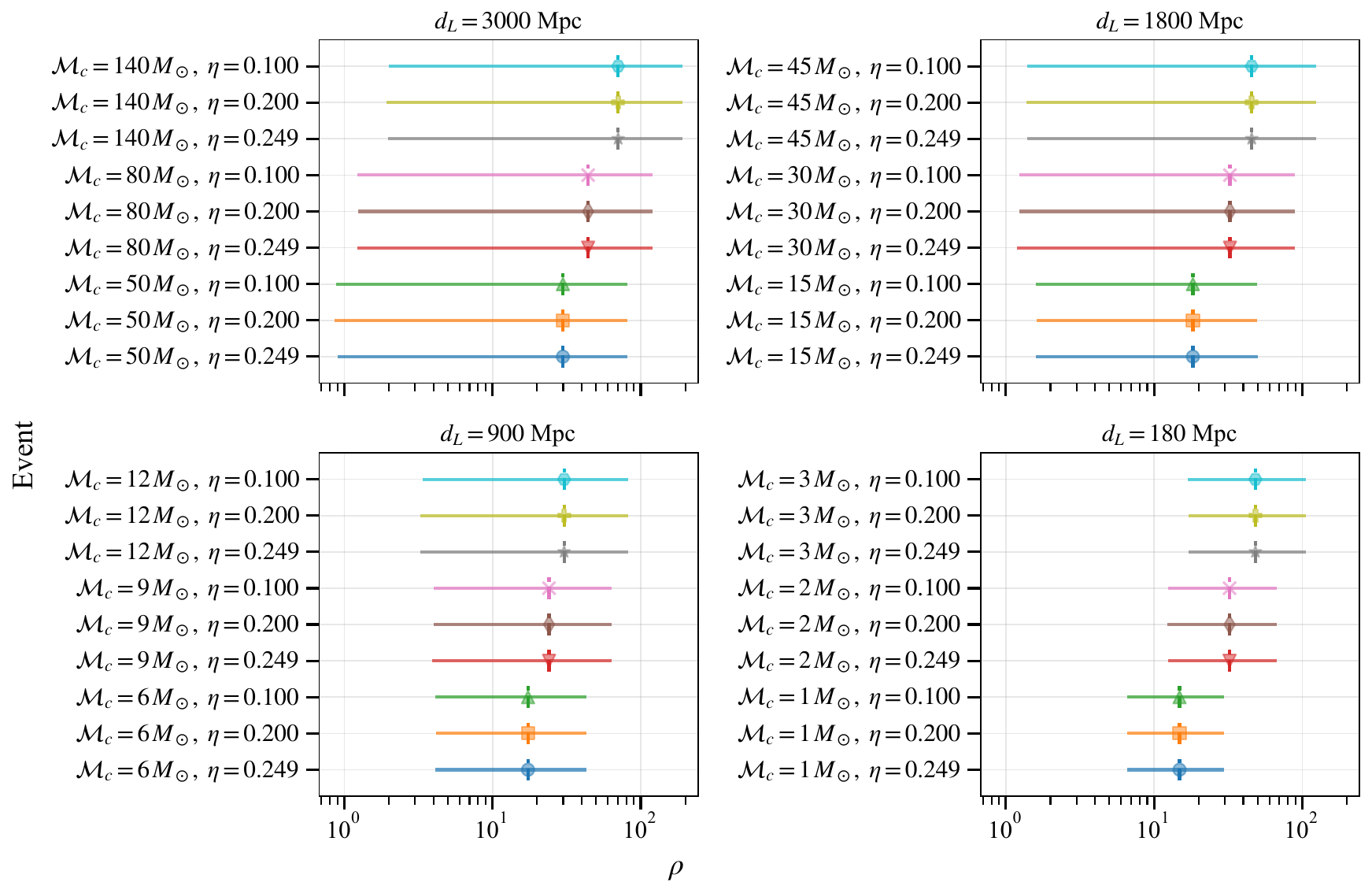}
    \caption{SNRs corresponding to the catalog used in Fig.~\ref{decigo-r1}. The horizontal lines indicate the range of SNRs for events across 1000 realizations with different angular parameters, and the markers denote the mean SNR averaged over these realizations.}
    \label{bd_snr}
\end{figure*}
\begin{figure*}[htbp]
    \centering
    \includegraphics[width=0.85\linewidth]{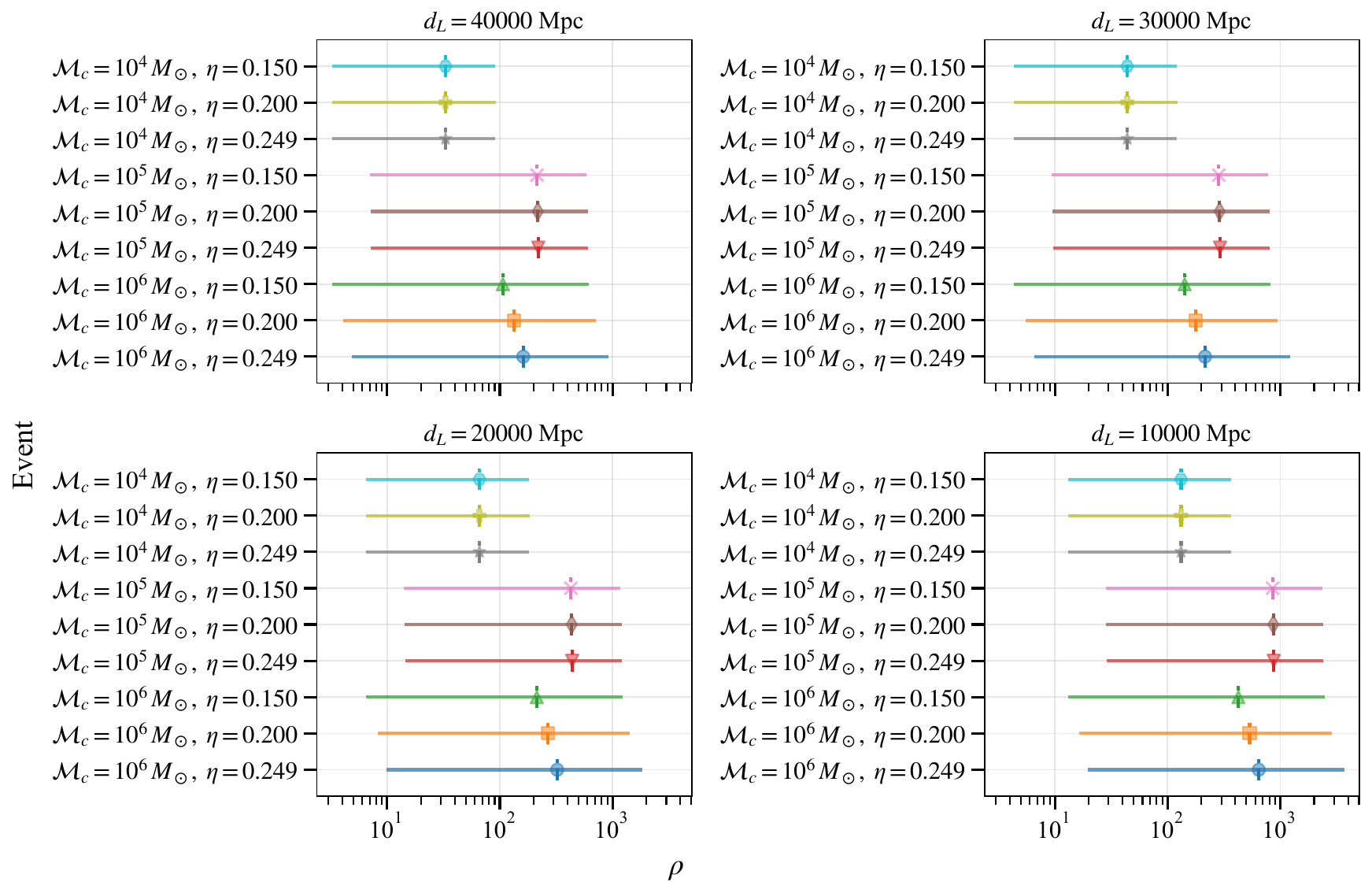}
    \caption{SNRs corresponding to the catalog used in Fig.~\ref{lisa-r1}. The horizontal lines indicate the range of SNRs for events across 1000 realizations with different angular parameters, and the markers denote the mean SNR averaged over these realizations.}
    \label{ls_snr}
\end{figure*}

\begin{figure*}[htbp]
  \centering
  \begin{minipage}[b]{0.49\linewidth}
      \centering
      \includegraphics[width=0.95\linewidth]{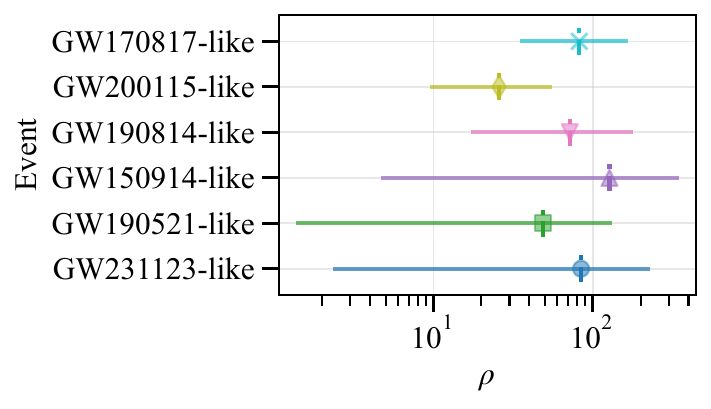}
      \caption{SNRs corresponding to the catalog used in Fig.~\ref{decigo1}. The horizontal lines indicate the range of SNRs for events across 1000
      realizations with different angular parameters, and the markers denote the mean SNR averaged over these realizations.}
      \label{fig:rho-bd}
  \end{minipage}
  \hfill
  \begin{minipage}[b]{0.49\linewidth}
      \centering
      \includegraphics[width=0.75\linewidth]{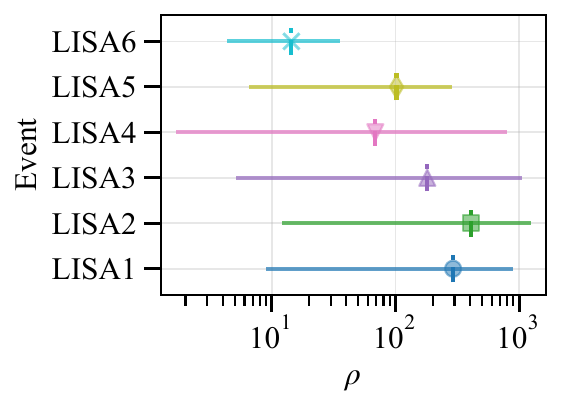}
      \caption{SNRs corresponding to the catalog used in Fig.~\ref{lisa1}. The horizontal lines indicate the range of SNRs for events across 1000
      realizations with different angular parameters, and the markers denote the mean SNR averaged over these realizations.}
      \label{fig:rho-ls}
  \end{minipage}
\end{figure*}

\begin{figure*}[htbp]
    \centering
    \includegraphics[width=0.8\linewidth]{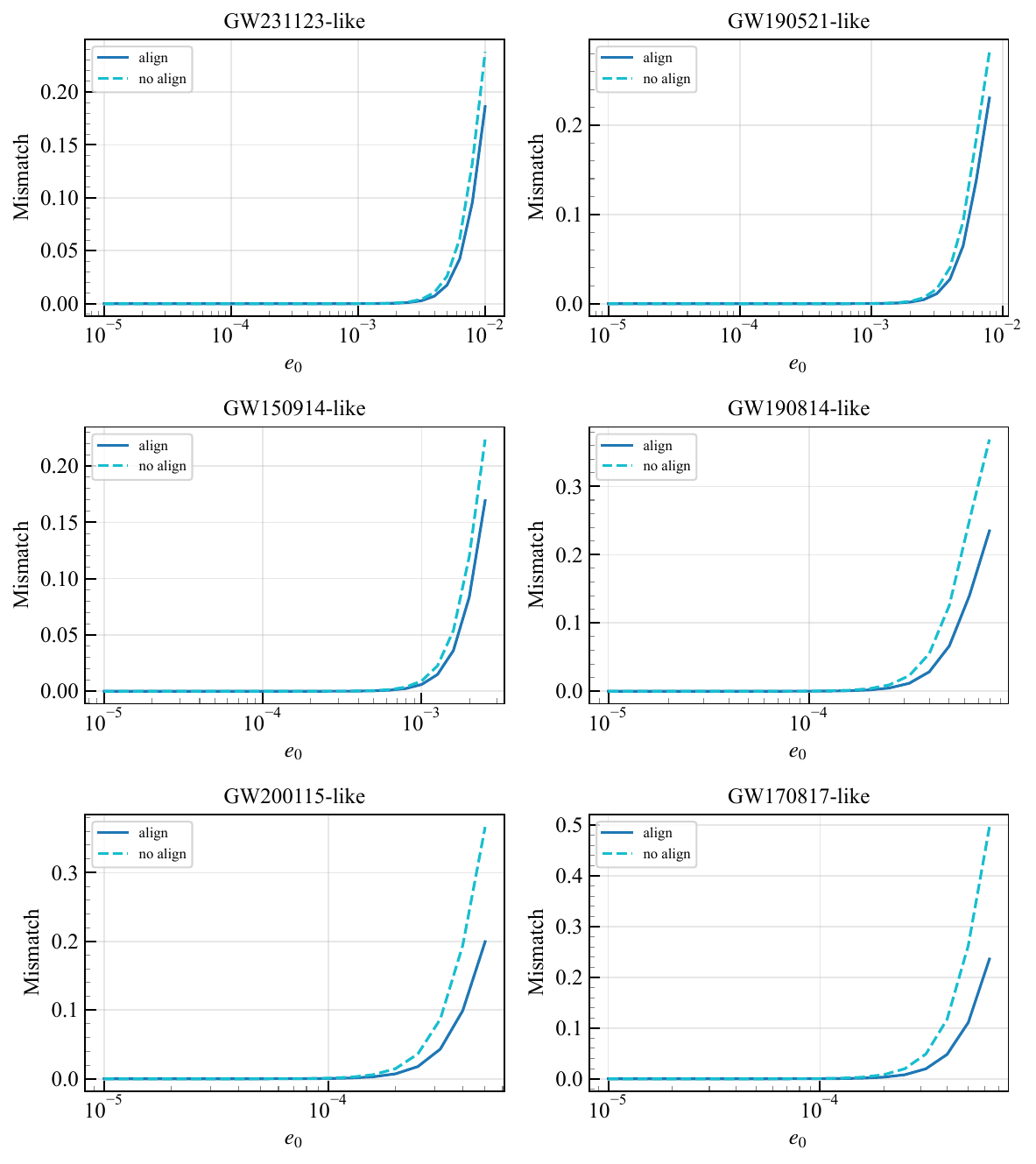}
    \caption{Comparison of waveform mismatches for B-DECIGO events with and without alignment. Alignment reduces the mismatch, thereby enhancing the validity for FCV estimates.}
    \label{mmnabd}
\end{figure*}
\begin{figure*}[htbp]
    \centering
    \includegraphics[width=0.8\linewidth]{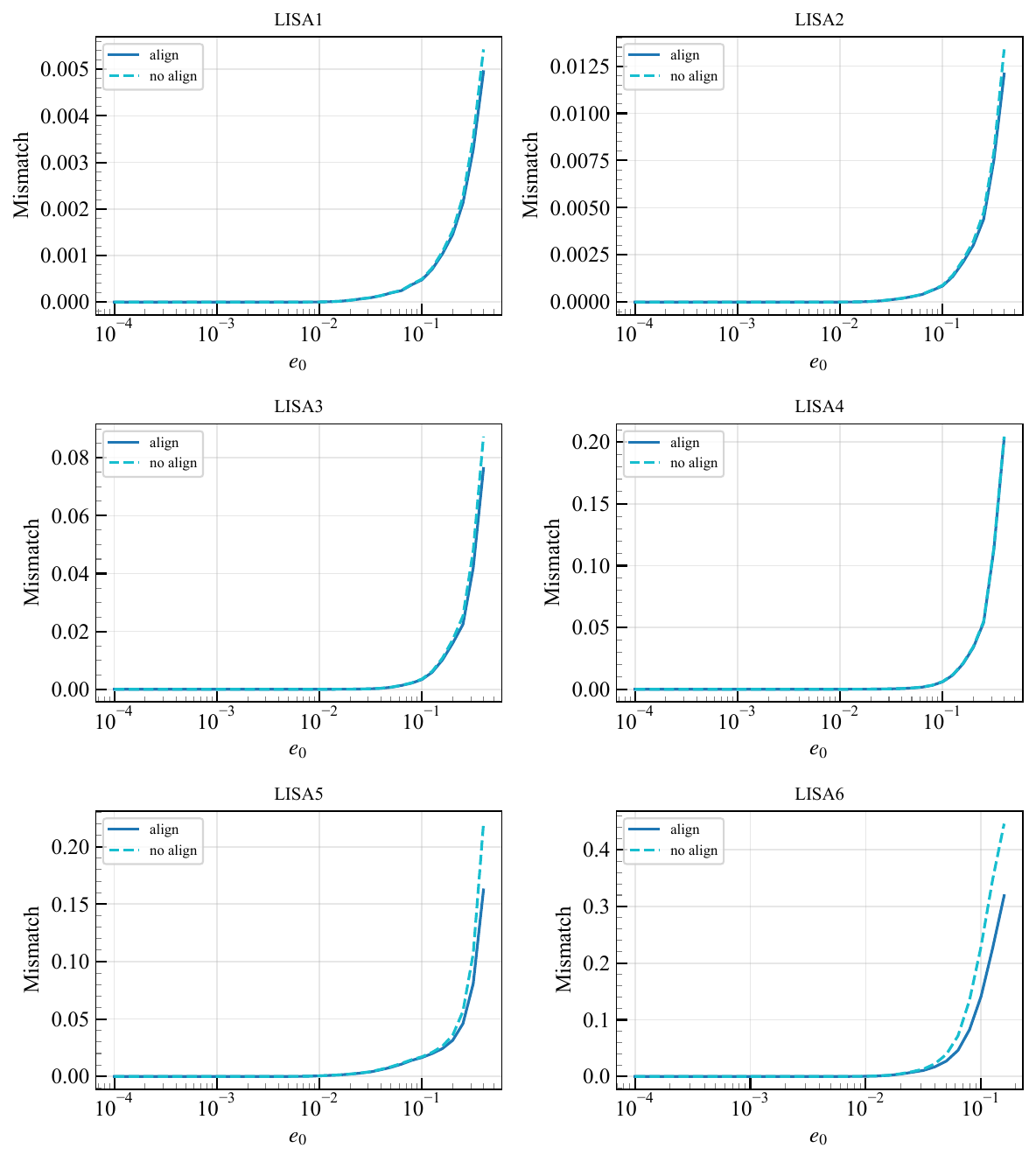}
    \caption{Comparison of waveform mismatches for LISA events with and without alignment. As observed for B-DECIGO, Alignment reduces the mismatch, thereby enhancing the validity of the FCV estimates.}
    \label{mmnals}
\end{figure*}

\section{\MakeUppercase{Systematic biases in a multi-detector network}}\label{appxfcv}

We consider a joint observation using a network of $N$ detectors. The best-fit (bf) parameter set $\vec\lambda^{\text{bf}}$ can be obtained by solving the generalized multidetector equation, which extends the single-detector maximum-likelihood condition to the network
\begin{eqnarray} \label{mll1}
	\sum_{d=1}^N \Big( s^d - h^{d,\text{AP}}(\vec\lambda^{\text{bf}}) \Big | \partial_{\lambda_i} h^d (\vec\lambda^{\text{bf}}) \Big) = 0\,,
\end{eqnarray}
where the superscript $d$ labels the $d$th detector, and $\partial_{\lambda_i}$ denotes the partial derivative with respect to the $i$th parameter in $\vec\lambda$. Both the waveform $h$ and its derivatives are evaluated at $\vec\lambda^{\text{bf}}$.
Since the detector strain $s^d$ contains both the signal and noise, $s^d = h^{d,s}(\vec\lambda^{\text{tr}}) + n^d$, Eq.~(\ref{mll1}) can be rewritten as
\begin{eqnarray}\label{mll2}
	&& \sum_{d=1}^N \Big( h^{d,s} (\vec\lambda^{\text{tr}}) - h^{d,\text{AP}}(\vec\lambda^{\text{bf}}) \Big | \partial_{\lambda_i} h^{d,\text{AP}} (\vec\lambda^{\text{bf}}) \Big) \nn
	&=& - \sum_{d=1}^N \Big( n^d \Big | \partial_{\lambda_i} h^{d,\text{AP}} (\vec\lambda^{\text{bf}}) \Big)\,.
\end{eqnarray}
We define the following quantity for the detector network,
\begin{equation}
	V^N_i (\vec\lambda^{\text{bf}}) = \sum_{d=1}^N \Big( n^d \Big |\partial_{\lambda_i} h^{d,\text{AP}} (\vec\lambda^{\text{bf}}) \Big)\,,
\end{equation}
then the FIM for the $N$-detector network can be written as,
\begin{equation}\label{fmx1}
	\Gamma^N_{ij} = \langle V^N_i V^N_j \rangle\,,
\end{equation}
where $\langle \cdot \rangle$ denotes the ensemble average. Using the fact that noise realizations are uncorrelated between detectors
\begin{equation}
	\langle (\tilde n^{l})^\ast(f')\ \tilde n^{m}(f) \rangle = \frac{1}{2} \text{Sn}(f) \delta_{lm} \delta(f-f')\,,
\end{equation}
Equation~(\ref{fmx1}) reduces to,
\begin{eqnarray}
	\Gamma^N_{ij} (\vec\lambda^{\text{bf}}) = \sum_{d=1}^N \Big( \partial_{\lambda_i} h^{d,\text{AP}} (\vec\lambda^{\text{bf}}) \Big| \partial _{\lambda_j} h^{d,\text{AP}} (\vec\lambda^{\text{bf}}) \Big)\,.
\end{eqnarray}
For a single detector, the FIM is
\begin{equation}
	\Gamma^d_{ij} = \langle v^d_i v^d_j \rangle = \Big( \partial_{\lambda_i} h^{d,\text{AP}} \Big| \partial _{\lambda_j} h^{d,\text{AP}} \Big)\,,
\end{equation}
with
\begin{equation}
	v^d_i  = \Big( n^d \Big | \partial_{\lambda_i} h^{d,\text{AP}} \Big)\,.
\end{equation}
Hence, the network FIM is simply the sum of the individual FIMs
\begin{equation}
    \Gamma^N_{ij} = \sum_{d=1}^N \Gamma^d_{ij}\,.
\end{equation}

The parameter measurement errors are defined as
\begin{eqnarray}
	\delta \lambda_i = \lambda^{\text{bf}}_i - \lambda^{\text{tr}}_i\,,
\end{eqnarray}
and the waveform error arising from using an approximate waveform $h^{\text{AP}}$ is
\begin{equation}\label{template_error}
	\delta h(\vec\lambda) = h^s(\vec\lambda) - h^{\text{AP}}(\vec\lambda)\,.
\end{equation}
Expanding the true waveform $h^s(\vec\lambda^{\text{tr}})$ to linear order in both $\delta \vec\lambda$ and $\delta h$ yields
\begin{eqnarray}\label{expan1}
	h^s(\vec\lambda^{\text{tr}}) &\approx& h^{\text{AP}}(\vec\lambda^{\text{bf}}) + \delta h(\vec\lambda^{\text{tr}}) - \delta \lambda_i \partial_{\lambda_i} h^{\text{AP}}(\vec\lambda^{\text{bf}}) \nn
	&&\quad + \mathcal O(2)\,,
\end{eqnarray}
where, for a network, $\delta \vec\lambda$ represents the overall parameter error from the joint observation. Substituting Eq.~(\ref{expan1}) into Eq.~(\ref{mll2}) gives 
\begin{eqnarray}
	&& \sum_{d=1}^N \Big( \delta h^d(\vec\lambda^{\text{tr}}) - \delta \lambda_i \partial_{\lambda_i} h^{d,\text{AP}}(\vec\lambda^{\text{bf}}) \Big | \partial_{\lambda_i} h^{d,AP} (\vec\lambda^{\text{bf}}) \Big) \nn
	&=& - \sum_{d=1}^N \Big( n^d \Big | \partial_{\lambda_i} h^{d,AP} (\vec\lambda^{\text{bf}}) \Big)\,.
\end{eqnarray}
At linear order, we approximate $\partial_{\lambda_i}h (\vec\lambda^{\text{tr}}) \approx \partial_{\lambda_i}h(\vec\lambda^{\text{bf}})$ \cite{Chandramouli:2024vhw}. The total bias for the network is
\begin{eqnarray}\label{bias_detn}
    \delta \lambda^{N}_i &=& \left(\Gamma^N(\vec\lambda^{\text{tr}})\right)^{-1}_{ij} \sum_{d=1}^N \Big( n^d \Big | \partial_{\lambda_j} h^{d,AP} (\vec\lambda^{\text{tr}}) \Big) +\nn
    && \left(\Gamma^N(\vec\lambda^{\text{tr}})\right)^{-1}_{ij} \sum_{d=1}^N \Big(\delta h(\vec\lambda^{\text{tr}}) \Big | \partial_{\lambda_j} h^{d,AP} (\vec\lambda^{\text{tr}}) \Big) \nn
    &=& \Delta \lambda_i^{N,\text{noise}} + \Delta \lambda_i^{N,\text{sys}}\,.
\end{eqnarray}
Equation~(\ref{bias_detn}) shows two contributions to the total error: one from detector noise, $\Delta \lambda_i^{N,\text{noise}}$, and one from systematic bias due to waveform inaccuracies, $\Delta \lambda_i^{N,\text{sys}}$.
For a signal measured by $N$ nearly identical detectors (i.e., similar locations and noise characteristics), both the FIM and the inner-product term scale with $N$. Consequently, the systematic bias does not change, consistent with \cite{Cutler:2007mi}. Similar derivations can also be found in Appendix A of \cite{Kapil:2024zdn}.

\section{\MakeUppercase{SNRs in the implemented catalogs}} \label{appx:rho}

We present the corresponding SNRs for the events considered in Secs.~\ref{sec:fcv-bdecigo} and \ref{sec:fcv-lisa}.
Figure~\ref{bd_snr} corresponds to the catalog used in Fig.~\ref{decigo-r1}, and Fig.~\ref{fig:rho-bd} corresponds to the catalog used in Fig.~\ref{decigo1}.
Similarly, Fig.~\ref{ls_snr} corresponds to the catalog used in Fig.~\ref{lisa-r1}, and Fig.~\ref{fig:rho-ls} corresponds to the catalog used in Fig.~\ref{lisa1}.

\section{\MakeUppercase{Waveform Alignment}} \label{appx:wa}

To assess whether waveform alignment extends the validity range of the FCV method or equivalently, whether it systematically reduces the mismatch, we compare mismatch results obtained with and without alignment. For B-DECIGO events, the comparison is shown in Fig.~\ref{mmnabd}, and the corresponding results for LISA events are presented in Fig.~\ref{mmnals}.

In all cases considered, waveform alignment leads to a systematic reduction in mismatch at larger values of $e_0$. This indicates that alignment mitigates differences between the injected and recovery waveforms, thereby extending the regime in which the linear-signal approximation remains valid for both B-DECIGO and LISA.

\FloatBarrier
\bibliographystyle{apsrev4-1}
\bibliography{main}

\end{document}